\documentclass[prd,showpacs,showkeys,nofootinbib,twocolumn,superscriptaddress]{revtex4-1}

\usepackage{graphicx} 
\usepackage{amsmath} 
\usepackage{amssymb} 
\usepackage{dsfont}

\begin{document}

\newcommand{\um}{\underline{m}}
\newcommand{\bS}{\bar{S}}
\newcommand{\br}{\bar{r}}
\newcommand{\bJ}{\bar{J}}
\newcommand{\bE}{\bar{E}}
\newcommand{\ba}{\bar{a}}

\title{Motion of spinning test bodies in Kerr spacetime} 

\author{Eva Hackmann}
\email{eva.hackmann@zarm.uni-bremen.de}
\affiliation{ZARM, University of Bremen, Am Fallturm, 28359 Bremen, Germany}

\author{Claus L\"ammerzahl}
\email{claus.laemmerzahl@zarm.uni-bremen.de}
\affiliation{ZARM, University of Bremen, Am Fallturm, 28359 Bremen, Germany}
\affiliation{Institut f\"ur Physik, Universit\"at Oldenburg, 26111 Oldenburg, Germany}

\author{Yuri N. Obukhov}
\email{obukhov@ibrae.ac.ru}
\affiliation{Theoretical Physics Laboratory, Nuclear Safety Institute, 
Russian Academy of Sciences, B.Tulskaya 52, 115191 Moscow, Russia} 

\author{Dirk Puetzfeld}
\email{dirk.puetzfeld@zarm.uni-bremen.de}
\homepage{http://puetzfeld.org}
\affiliation{ZARM, University of Bremen, Am Fallturm, 28359 Bremen, Germany}

\author{Isabell Schaffer}
\email{trisax@t-online.de}
\affiliation{ZARM, University of Bremen, Am Fallturm, 28359 Bremen, Germany}
\affiliation{Institut f\"ur Physik, Universit\"at Oldenburg, 26111 Oldenburg, Germany}

\date{ \today}

\begin{abstract}
We investigate the motion of spinning test bodies in General Relativity. By means of a multipolar approximation method for extended test bodies we derive the equations of motion, and classify the orbital motion of pole-dipole test bodies in the equatorial plane of the Kerr geometry. An exact expression for the periastron shift of a spinning test body is given. Implications of test body spin corrections are studied and compared with the results obtained by means of other approximation schemes.  
\end{abstract}

\pacs{04.25.-g; 04.20.-q}
\keywords{Approximation methods; Equations of motion; Extreme mass ratios}

\maketitle


\section{Introduction}\label{introduction_sec}

Extreme mass ratios in astrophysical situations, for example as found in the galactic center, allow for an approximate analytic description of the motion in certain parameter regimes. The steadily improving observational situation of the galactic center \cite{Gillessenetal2009,Genzel:etal:2010,SKA:Web} may soon enable us to test different competing theoretical approaches to model the motion of astrophysical objects in the theory of General Relativity.   

In this work we study the motion of extended spinning test bodies in a Kerr background. Our starting point is an explicit velocity formula based on the multipolar description \cite{Mathisson:1937,Papapetrou:1951:3,Tulczyjew:1959,Dixon:1964,Dixon:1974:1} of pole-dipole test bodies, with the help of which we classify the orbital motion in the equatorial plane of a Kerr black hole for aligned and anti-aligned test body spin. An exact expression for the periastron shift is given and compared with corresponding post-Newtonian results. We provide an estimate of the test body spin corrections for orbits around the black hole in the galactic center.

The structure of the paper is as follows. In section \ref{eom_section} we provide the equations of motion for spinning test bodies and derive a general formula which relates the momentum and the velocity of the test body. The motion of spinning test bodies is then studied in a Kerr background in section \ref{orbits_sec}. These equations of motion are of a mathematical structure which allows for an analytic solution \cite{EHKKL11,EHKKLS12} and a systematic classification of different orbit types in section \ref{classifcation_sec}. In section \ref{perihel_sec} a general formula for the periastron shift is given and compared to corresponding post-Newtonian results. Our conclusions are drawn in section \ref{conclusions_sec}. In the appendices \ref{dimension_acronyms_app} and \ref{app:FD} we provide some supplementary material and a summary of our conventions.

\section{Equations of motion of spinning test bodies}\label{eom_section}

The equations of motion of spinning extended test bodies up to the pole-dipole order have been derived in several works \cite{Mathisson:1937,Papapetrou:1951:3,Tulczyjew:1959,Dixon:1964,Dixon:1974:1,Steinhoff:Puetzfeld:2009:1} by means of different multipolar approximation techniques and are given by the following set of equations:
\begin{eqnarray}
\frac{D p_a }{d s} &=& \frac{1}{2} R_{abcd} u^{b} S^{cd},  \label{pap1}\\
\frac{D S^{ab}}{d s} &=& 2 p^{[a} u^{b]} . \label{pap2}
\end{eqnarray}
Here $u^a:=dY^a/ds$ denotes the 4-velocity of the body along its world line (normalized to $u^a u_a=1$), $p^a$ the momentum, $S^{ab} = -S^{ba}$ the spin, $\frac{D}{ds}$ the covariant derivative along $u^a$, and $R_{abcd}$ is the Riemannian curvature. Eq.\ \eqref{pap2} implies that the momentum is given by
\begin{equation}
p^a = m u^a + \frac{D S^{ab}}{d s} u_b, \label{generalized_momentum}
\end{equation}
where $m := p_a u^a$. Note that in order to close the system of equations (\ref{pap1})--(\ref{pap2}) a supplementary condition has to be imposed.

\subsection{Conserved quantities}
If $\xi^a$ is a Killing-vector, i.e.\ $\nabla_{(b} \xi_{a)} = 0$, then the quantity
\begin{equation}
E_{\xi} = p_a \xi^a + \frac{1}{2} S^{ab} \nabla_a \xi_b ,
\end{equation}
is conserved, see e.g.~\cite{Ehlers:Rudolph:1977,Steinhoff:Puetzfeld:2009:1} for a derivation. 

Other conserved quantities depend on the supplementary condition. In the pole-dipole case, the spin length $S$ given by 
\begin{equation}
S^2 := \tfrac12 S_{ab} S^{ab} \label{DefSpinLength}
\end{equation} 
is conserved for the two well-known supplementary conditions of Tulczyjew 
\begin{equation}
p_a S^{ab} = 0\,,\label{TulczyjewCondition}
\end{equation} 
and Frenkel 
\begin{equation}
u_a S^{ab} = 0 \,. \label{FrenkelCondition} 
\end{equation}

Apart from $m$ one may define a mass $\underline{m}$ by $\underline{m}^2 := p^a p_a$. In the pole-dipole case $\underline{m}$ is conserved if one chooses Tulczyjew's spin supplementary condition \eqref{TulczyjewCondition}. However, for the Frenkel condition \eqref{FrenkelCondition} the mass $m$ is conserved in the pole-dipole case. 

\subsection{Velocity--momentum relation}

For either of the two supplementary conditions (\ref{TulczyjewCondition}) or (\ref{FrenkelCondition}) the following relation, see \cite{Obukhov:Puetzfeld:2011:1} for a derivation, between the velocity and the momentum holds:
\begin{eqnarray}
u^a \stackrel{(\ref{TulczyjewCondition}) \, \vee  \, (\ref{FrenkelCondition})}{= \,\,} \hat{p}^a + \frac{2 S^{ac} S^{de} R_{decb}}{4 \underline{m}^2 + S^{cd} S^{ef} R_{cdef}} \hat{p}^b \label{mom_vel_relation},
\end{eqnarray}
where 
\begin{eqnarray}
\hat{p}^a &:=& \frac{m}{\underline{m}^2} p^a - \frac{1}{\underline{m}^2} \frac{D}{ds} \left(S^{ab} p_b \right). \label{def_hat_mom} 
\end{eqnarray}
From the velocity formula (\ref{mom_vel_relation}) and the normalization condition $u_a u^a=1$, we obtain -- for the Tulczyjew condition (\ref{TulczyjewCondition}) -- an explicit expression for the mass $m$ of the following form
\begin{eqnarray} 
m \stackrel{(\ref{TulczyjewCondition})}{=}{\frac{\underline{m}^2}{\mu}}, \label{m_rewritten_tul}
\end{eqnarray}
where we introduced auxiliary quantities
\begin{eqnarray} 
\mu^2 &=& \underline{m}^2 + A_1^2 S^{ab} S^{cd} R_{cdbe} p^e S_{af} S_{gh} R^{ghfi} p_i,\label{mu}\\
A_1 &=& \frac{2}{4 \underline{m}^2 + S^{ab} S^{cd} R_{cdab}}\,.
\end{eqnarray}
In flat spacetime we have $m= \mu = \underline{m}$. 
Resubstituting (\ref{m_rewritten_tul}) back into (\ref{mom_vel_relation}) we obtain an expression for the velocity
\begin{eqnarray}
u^a  &\stackrel{(\ref{TulczyjewCondition})}{=}&  K^a{}_b\,p^b,\label{vel_final_formula}\\
K^a{}_b &=& {\frac{1}{\mu}}\left(\delta^a_b - A_1 S^{ae} S^{cd} R_{becd}\right),\label{Kab}
\end{eqnarray}
as a function of the momentum, the mass $\underline{m}$, and the spin, i.e.\ $u^a=u^a\left(\underline{m},p^a,S^{ab},R_{abcd} \right)$.

\subsection{Nonlinear dynamics of spin}

For the Tulczyjew condition (\ref{TulczyjewCondition}) we define the spin vector $S^a$ as 
\begin{eqnarray}
S^a&:=& \frac{1}{2 \underline{m} \sqrt{-g}} \varepsilon^{abcd} p_b S_{cd}, \quad  S^{ab}= \frac{1}{\underline{m} \sqrt{-g}} \varepsilon^{abcd} p_c S_{d}.\nonumber \\
 \label{S_vector}
\end{eqnarray}
Here $\varepsilon^{abcd}$ is the totally antisymmetric Levi-Civita symbol; the only nontrivial component is equal $\varepsilon^{0123}  = 1$. It is straightforward to derive the equation of motion for the vector of spin:
\begin{equation}
{\frac{D S^a }{d s}} = {\frac {p^ap_b}{\underline{m}^2}}{\frac{D S^b }{d s}}.\label{vector_dynamics1}
\end{equation}
By construction, we have the orthogonality 
\begin{equation}
p_aS^a = 0,\label{pS0}
\end{equation}
and noticing that $S^{ab}S_b = 0$ from (\ref{S_vector}), we use (\ref{vel_final_formula}) and (\ref{Kab}) to verify another orthogonality property
\begin{equation}
u_aS^a = 0,\label{uS0}
\end{equation}
Substituting the velocity-momentum relation (\ref{vel_final_formula}) and (\ref{Kab}) into (\ref{pap1}) and (\ref{vector_dynamics1}), we derive the closed system of dynamical equations for the momentum and spin vectors 
\begin{eqnarray}
\frac{D p_a }{d s} &=& {\frac{1}{2}}K^e{}_a R_{ebcd} p^{b} S^{cd},  \label{pap1n}\\
\frac{D S^{a}}{d s} &=& -\,{\frac{p^ap^b}{2\underline{m}^2}} K^e{}_f R_{ebcd}S^{cd}S^f. \label{pap2n}
\end{eqnarray}
Contracting (\ref{vector_dynamics1}) or (\ref{pap2n}) with $S_a$, we verify that the length of the spin vector is constant by making use of (\ref{pS0}). From (\ref{S_vector}) we find 
\begin{equation}
S_a S^a = - {\frac 12}S_{ab} S^{ab} = - S^2,\label{S2} 
\end{equation}
where we recall the definition of the spin length \eqref{DefSpinLength}. The vector of spin is therefore spacelike. 

The dynamical equations (\ref{pap1n}) and (\ref{pap2n}) are highly nonlinear in spin. Indeed, the right-hand sides of these equations contain 
\begin{eqnarray}
K^a{}_b &=& {\frac 1{\mu}}\Biggl[\delta^a_b - 2A_1\left(\delta^a_c - {\frac {p^ap_c}{\underline{m}^2}}\right)(S^2\delta^c_d + S^cS_d) \nonumber\\
&&  \times \left\{R^d{}_b + R^d{}_{kbn}\left({\frac {p^kp^n}{\underline{m}^2}} - {\frac {S^kS^n}{S^2}}\right)\right\}\Biggr].\label{Kab2}
\end{eqnarray}
Furthermore, we explicitly have
\begin{eqnarray}
\mu^2 &=& \underline{m}^2 + 4A_1^2\left[S^2(R_{ab}p^aS^b)^2\right.\nonumber\\
&& -\,{\frac 1{\underline{m}^2}}(S^2R_{ab}p^ap^b - R_{acbd}p^ap^bS^cS^d)^2\nonumber\\
&& +\, (S^2R^{ab} - R^{aebf}S_eS_f)p_b \nonumber\\
&& \left. \times (S^2R_{ac} - R_{akcn}S^kS^n)p^c \right],\label{mu2}\\
{\frac 1{A_1}} &=& 2\left[\underline{m}^2 - R_{acbd}S^aS^bp^cp^d/\underline{m}^2\right.\nonumber\\
&& \left. -\,R_{ab}(S^aS^b + g^{ab}S^2/2 + p^ap^b/\underline{m}^2)\right].\label{A1a}
\end{eqnarray}
Here $R_{ab}$ is the Ricci tensor. 

The analysis of the nonlinear system (\ref{pap1n}) and (\ref{pap2n}) is a complicated problem, in general. A perturbation scheme was developed in \cite{Chicone:2005,Mashhoon:2006,Singh:2008a,Singh:2008b} to deal with the full nonlinear system. In this approach, one linearizes the equations of motion to obtain
\begin{eqnarray}
\frac{D p_a }{d s} &\approx& {\frac{1}{2}}R_{abcd} p^{b} S^{cd},  \label{pap1lin}\\
\frac{D S^{a}}{d s} &\approx& 0, \label{pap2lin}
\end{eqnarray}
and the solution of the full system is then constructed as a series in the powers of spin $S$ which is used as a perturbation parameter. In the linearized case, we again have $m= \mu = \underline{m}$ and hence $p^a \approx \underline{m}u^a$. It is worthwhile to note that the Gravity Probe B experiment \cite{Schiff:1960,GPB:2011} is actually based on the linearized equations of motion (\ref{pap1lin}) and (\ref{pap2lin}).

In this paper, we analyze the complete nonlinear equations of motion without using approximations and perturbation theory. 

\section{Equations of motion in a Kerr background}\label{orbits_sec}

In the following, we are going to study test bodies endowed with spin in the gravitational field of a rotating source described by the Kerr metric. This problem was investigated in the past for the Tulczyjew supplementary condition (see \cite{Semerak:1999,Singh:2008b}, e.g.), as well as for the Frenkel condition \cite{Semerak:2007,Plyatsko:2013}. In view of the complexity of the problem, the solution in most cases was obtained numerically and/or approximately with the help of perturbation theory. 

Here we will specialize to the integrable case for which we obtain an exact and analytical result. The full nonlinear equations of motion are considered, no linearization or other approximation is made. Since the Kerr metric satisfies the vacuum Einstein field equation, $R_{ab} = 0$, the formulas (\ref{Kab2})-(\ref{A1a}) become significantly simpler. 

\subsection{The Kerr metric}

In Boyer-Lindquist coordinates $(t,r,\theta,\phi)$, the Kerr metric takes the form
\begin{align}
ds^2&= \left(1-\frac{2Mr}{\rho^2}\right)dt^2+\frac{4aMr {\rm sin}^2\theta}{\rho^2} dtd\phi - \frac{\rho^2}{\Delta} dr^2\nonumber\\
& \quad - \rho^2d\theta^2 -{\rm sin}^2\theta \left(r^2+a^2+\frac{2a^2Mr {\rm sin}^2\theta}{\rho^2} \right)d\phi^2\,, \label{kerr_metric}
\end{align}
where $M$ is the mass parameter, $a$ the Kerr parameter, and
\begin{align}
\Delta & := r^2 - 2 M r + a^2\,, \\ 
\rho^2 & := r^2 + a^2 {\rm cos}^2\theta \,.
\end{align}
The Kerr metric allows for two Killing vector fields given by:
\begin{eqnarray}
{\underset{_E}\xi^a}=\delta^a_t, \quad \quad {\underset{_J}\xi^a}=\delta^a_\phi. \label{kerr_killing}
\end{eqnarray}
Furthermore, we have
\begin{eqnarray}
\sqrt{-g}:=\sqrt{-{\rm det}\left(g_{ab}\right)}=\rho^2 {\rm sin}\theta. 
\end{eqnarray}

\subsection{Equatorial orbits for polar spin}\label{sub_eq_orbits}

Let us assume that the spin vector of a test body has only one, namely polar, component:
\begin{equation}
S^a = S^\theta \delta^a_\theta.\label{S_theta}
\end{equation}
In view of the orthogonality relations (\ref{pS0}) and (\ref{uS0}) the polar ansatz (\ref{S_theta}) yields
\begin{eqnarray}
p^{\theta}=0,\qquad u^{\theta}=0. \label{up0}
\end{eqnarray} 
Recalling $u^{\theta} = d\theta/ds$, we thus conclude that the polar angle is fixed, $\theta =$ const. Therefore, we can focus on equatorial orbits, i.e. 
\begin{eqnarray}
\theta = \frac{\pi}{2}. \label{eq_orbits_def}
\end{eqnarray}
The consistency of the equatorial setup (\ref{S_theta})-(\ref{eq_orbits_def}) was analyzed earlier in \cite{Steinhoff:Puetzfeld:2012}. It is worthwhile to note that the assumption (\ref{S_theta}) on the equatorial plane means that the spin of a test body is aligned with the spin of the Kerr source. 

Let us now turn to the integration of the equations of motion (\ref{pap1n}) and (\ref{pap2n}). The polar ansatz (\ref{S_theta}) and its corollary (\ref{up0}) leave us with the four unknowns $\{p^t,p^r,p^\phi,S^{\theta}\}$, which should be determined from the equations of motion. Fortunately, we have exactly four integrals of motion and we can find the nontrivial components of the vectors of momentum and spin from the following set of equations
\begin{align}
S^2 & = -\,S_a S^a,\label{Sint}\\
\underline{m}^2 & = p_a p^a, \label{pint} \\
E &= p_a \Bigl( {\underset{_E}\xi^a} + {\frac{\varepsilon^{abcd}}{2\underline{m}\sqrt{-g}}} S_b \nabla_c {\underset{_E}\xi}{}_d\Bigr), \label{Eint}  \\
-J &= p_a \Bigl( {\underset{_J}\xi^a} + {\frac{\varepsilon^{abcd}}{2\underline{m}\sqrt{-g}}} S_b \nabla_c {\underset{_J}\xi}{}_d\Bigr), \label{Jint}
\end{align}
in terms of the mass $\underline{m}$, the spin length $S$, the energy $E$, and the angular momentum $J$. 

From the length conservation of spin (\ref{Sint}), we immediately find $S^{\theta} = S / \sqrt{- g_{\theta\theta}}$. For completeness we can use (\ref{S_vector}) to write down the nontrivial components of the spin tensor in the equatorial plane:
\begin{eqnarray}
S^{rt}&=&-\frac{S p_\phi}{\underline{m} r}, \quad S^{\phi t}=\frac{S p_r}{\underline{m} r}, \quad S^{\phi r}=-\frac{S p_t}{\underline{m} r}.
\end{eqnarray}
The algebraic system (\ref{Eint}) and (\ref{Jint}) can be solved for the momentum components $p_t$ and $p_\phi$ in terms of the constants of motion:
\begin{eqnarray}
p_t&=&\frac{E-\frac{MS}{\underline{m}r^3}\left(J-aE\right)}{1-\frac{MS^2}{\underline{m}^2r^3}}\,, \label{p_t_explicit} \\
p_\phi&=&\frac{-J-\frac{aMS}{\underline{m}r^3}\left[aE\left(1-\frac{r^3}{a^2M}\right)-J\right]}{1-\frac{MS^2}{\underline{m}^2r^3}} \label{p_phi_explicit}\,.
\end{eqnarray}
The remaining component $p_r$ is obtained from (\ref{pint}).

\subsection{Orbital equation of motion}

With the help of (\ref{vel_final_formula}), (\ref{Kab2})-(\ref{A1a}), (\ref{p_t_explicit}), and (\ref{p_phi_explicit}) we can derive explicit expressions for the velocity components in terms of the constants of motion and the parameters of the test body, i.e.\ $u^a=u^a\left(\underline{m},S,E,J,a,M\right)$.  From this we derive an explicit expression for $u^{\br}/u^\phi$,
\begin{equation}
\frac{d\br}{d\phi} = \frac{\bar{\Delta} (\br^3 + \bS^2)}{\br \bar Q} \sqrt{\bar P_{\ba}}\,, \label{Kdrdphinorm}
\end{equation}
where we introduced the dimensionless quantities
\begin{align}
\br&=\frac{r}{M}\,, \quad\ba=\frac{a}{M}\,, \quad\bJ=\frac{J}{\um M}\,,\nonumber\\
\bE&=\frac{E}{\um}\,, \quad\bS=\frac{S}{\um M}\,,\label{dimensionless}
\end{align}
and
\begin{widetext}
\begin{align}
\bar P_{\ba} & = (\bE^2-1)\br^8 + 2\br^7 + (\ba^2(\bE^2-1) - (\bS\bE-\bJ)^2)\br^6 + 2 ((\bS\bE-\bJ)^2+\bS(\bS-\bE\bJ)+\ba\bE(\ba\bE+3\bS\bE-2\bJ)) \br^5 \nonumber\\
& \quad - 4 \bS^2 \br^4 + 2\ba\bS (\bJ^2-\bE\bS\bJ+\ba^2\bE^2+\bE^2\bS\ba-2\ba\bJ\bE+\bS\ba) \br^3 + \bS^2 ((\ba\bE-\bJ)^2-\bS^2) \br^2 + 2 \bS^4 \br - \ba^2\bS^4, \label{defPa}\\
\bar Q & = (\bJ-\bS\bE)\br^6 + (\bS\bE+\ba\bE-\bJ)(2\br^5+\bS(\ba+2\bS)\br^3-4\bS^2\br^2+3\ba^2\bS^2\br)+\bS^3\ba(\ba\bE-\bJ),
\end{align}
\end{widetext}
with $\bar{\Delta}=\br^2-2\br+\ba^2$. In the following all quantities with a bar are always dimensionless. 

\subsection{Integration}\label{integration_sec}

The equation of motion \eqref{Kdrdphinorm} can be integrated analytically in a parametric form. First we notice that the corresponding integral equation
\begin{equation}
\phi-\phi_0 = \int_{\br_0}^{\br}  \frac{\br \bar{Q}}{\bar\Delta(\bS^2+\br^3)\sqrt{\bar P_{\ba}}}d\br\,,
\end{equation}
contains on the right hand side a hyperelliptic integral of genus three and the third kind. The corresponding problem for genus two was recently solved analytically \cite{Garciaetal2013} in a parametric form by introducing a new affine parameter $\lambda$, which may be considered as an analogue of the Mino time \cite{Mino03}. Together with the analytic solution of integral equations involving hyperelliptic integrals of genus three and the first kind \cite{EHKKL11}, the solution $\br(\lambda)$ and $\phi(\lambda)$ can be found analytically. However, we will not elaborate this here but rather focus on the related classification of the orbits, and on the periastron shift in section \ref{perihel_sec}. \medskip

\section{Classification of orbital motion}\label{classifcation_sec}

We will now analyze the orbital motion in the considered setting of equatorial motion with aligned spin. Observe that the substitutions $(\ba,\bJ,\bS)\to(-\ba,-\bJ,-\bS)$ and $(\bE,\bJ)\to(-\bE,-\bJ)$ only change the sign of the equation of motion, $\frac{d\br}{d\phi}\to-\frac{d\br}{d\phi}$. Therefore, this only reverses the direction but leaves the type of orbit unchanged, so we choose $\ba\geq0$ and $\bE\geq0$.

\subsection{Circular motion}

From equation \eqref{Kdrdphinorm} it can be inferred that the expression under the square root given by \eqref{defPa} has to be positive to get physical meaningful results.  Only if $\bar P_{\ba}\geq0$ motion is possible for the given parameters of the spacetime and the particle. The points $\bar P_{\ba}=0$ define the turning points of the motion. Coinciding turning points correspond to circular orbits and are given by double zeros of $\bar P_{\ba}$, 
\begin{align}
\bar P_{\ba}=0\,, \quad \quad  \frac{d\bar P_{\ba}}{d\br}=0\,.\label{Cond_doubles}
\end{align}
Solving this two conditions for $\bE$ and $\bJ$ yields
\begin{widetext}
\begin{align}
\bE_{1,2,3,4} & = \pm \frac{\sqrt{\br}}{\sqrt{2}\br^3\bar V_{\ba}} \Big\{ \bar V_{\ba} \big( \bar R_{\ba} \pm \bar{\Delta} [\bar U_{\ba}^2 (9\ba^2\bS^4+6\bS\br^2(\bS^2+2\br^3)\ba+\br(4\br^6+13\br^3\bS^2-8\bS^4))]^{\frac{1}{2}} \big) \Big\}^{\frac{1}{2}}\,, \label{CircE}\\
\bJ_{1,2,3,4} & = \frac{1}{\br^3\bE \bar U_{\ba}}\big\{ (3\br\bS^5+3\br^7(2\bE^2-1)\bS)\ba^3+(\bS^6+\br^3(\bE^2-3\br+6)\bS^4+3\br^6(\br\bE^2+\br-2+4\bE^2)\bS^2\nonumber\\
& \qquad +\br^{9}(2\bE^2-1))\ba^2+(\br^2(\br-4)\bS^5-\br^5(3\br^2\bE^2-9\bE^2-4\br-4\br\bE^2+4)\bS^3+\br^8(8\br\bE^2-5\br+8)\bS)a -\br\bS^6 \nonumber\\
& \qquad -\br^4(2\br-3)(\br+\br\bE^2-4)\bS^4+\br^7(\br^2+9-7\br+3\br\bE^2)\bS^2-\br^{10}(-3\br\bE^2+\br^2\bE^2+4\br-4-\br^2)\big\}\,, \label{CircJ}
\end{align}
where
\begin{align}
\bar V_{\ba} & =-6\bS\br(\bS^2+2\br^3)\bar{\Delta}\ba +((3\br-4)\bS^4+\br^3(6\br-19)\bS^2-4\br^6)\ba^2 +\br(2\br-3)^2\bS^4+\br^4(4\br^2-27\br+36)\bS^2+\br^7(\br-3)^2\,,\\
\bar R_{\ba} & = -18 \bS^3\ba^5\br^4 +(3\bS^6+3(3\br-4)\br^3\bS^4+18\bS^2\br^6)\ba^4 +(8\br^2\bS^5-2\br^5(27\br-67)\bS^3-2\br^8(9 \br-19)\bS)\ba^3\nonumber\\
& \quad +(5\br(\br-2)\bS^6+\br^4(15\br^2+87-70\br)\bS^4+\br^7(12\br^2+30-37\br)\bS^2-2\br^{10}(3\br-5))\ba^2\nonumber\\
& \quad -(4\br^3(\br^2+5-4\br)\bS^5+2\br^6(-77\br+14\br^2+94)\bS^3+2\br^9(11\br^2-41\br+40)\bS)\ba\nonumber\\
& \quad +2\br^2(3-2\br)\bS^6 +(8\br^3-56\br^2+145\br-126)\br^5\bS^4 +(8\br^3-65\br^2+160\br-126)\br^8\bS^2 +2\br^{11}(\br-3)(\br-2)^2,\\
\bar U_{\ba} & = 6\bS\br^4\ba^2+(\bS^4+12\br^3\bS^2-3\br^4\bS^2+2\br^6)\ba-\bS\br^2(4\bS^2\br-\br^4-9\bS^2)\,.
\end{align}
\end{widetext}

\begin{figure*}
\includegraphics[width=0.32\textwidth]{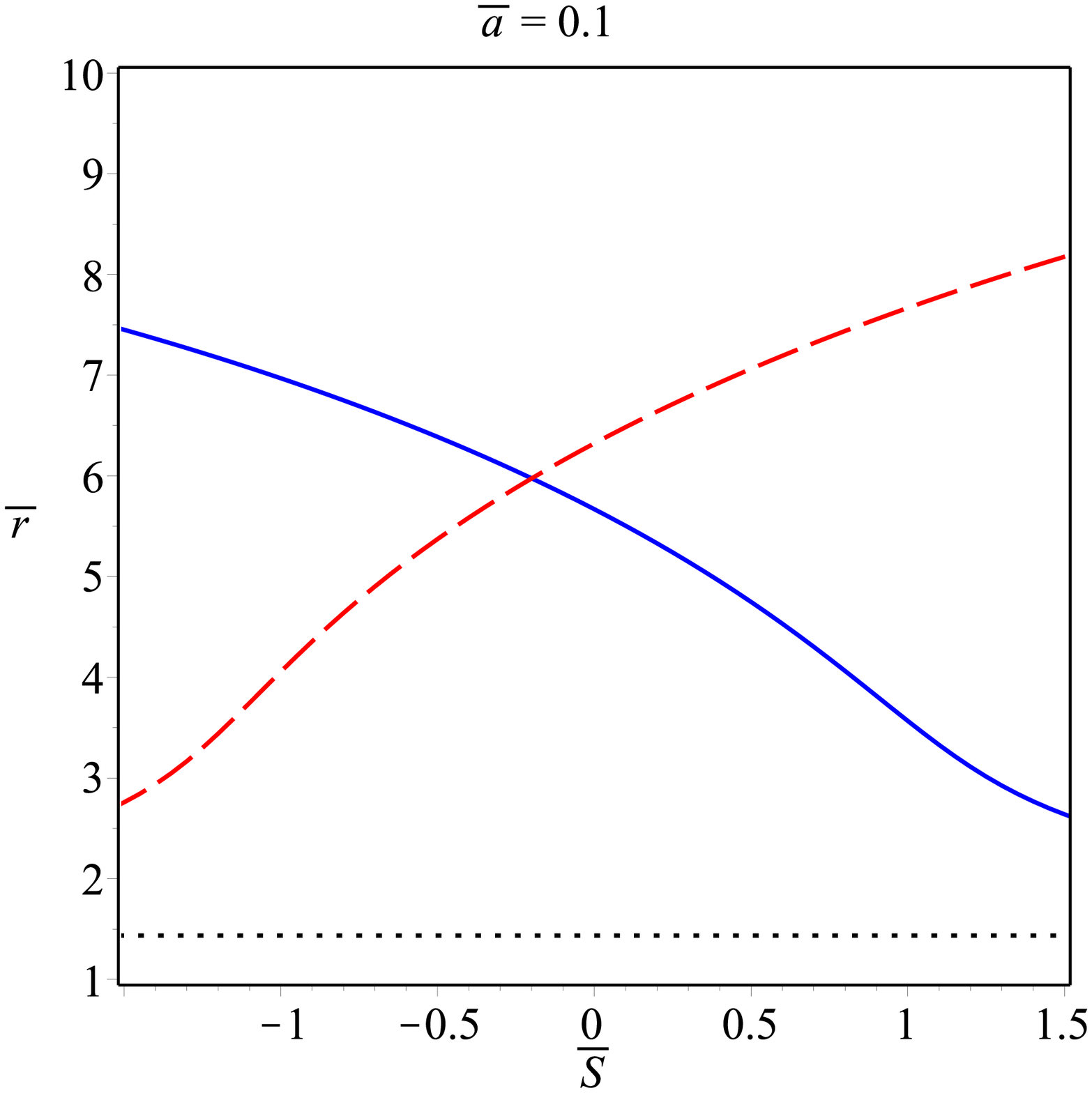}
\includegraphics[width=0.32\textwidth]{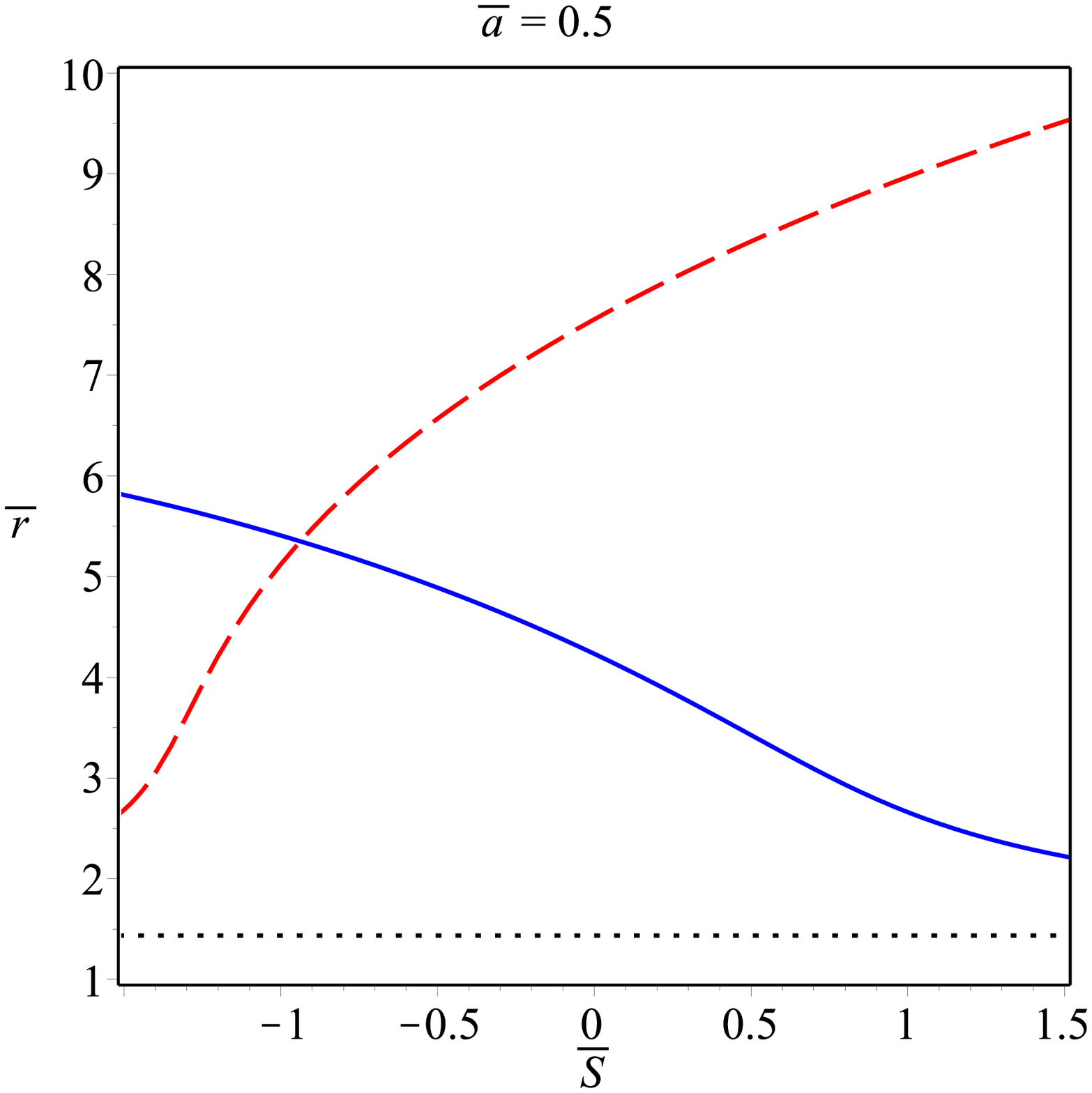}
\includegraphics[width=0.32\textwidth]{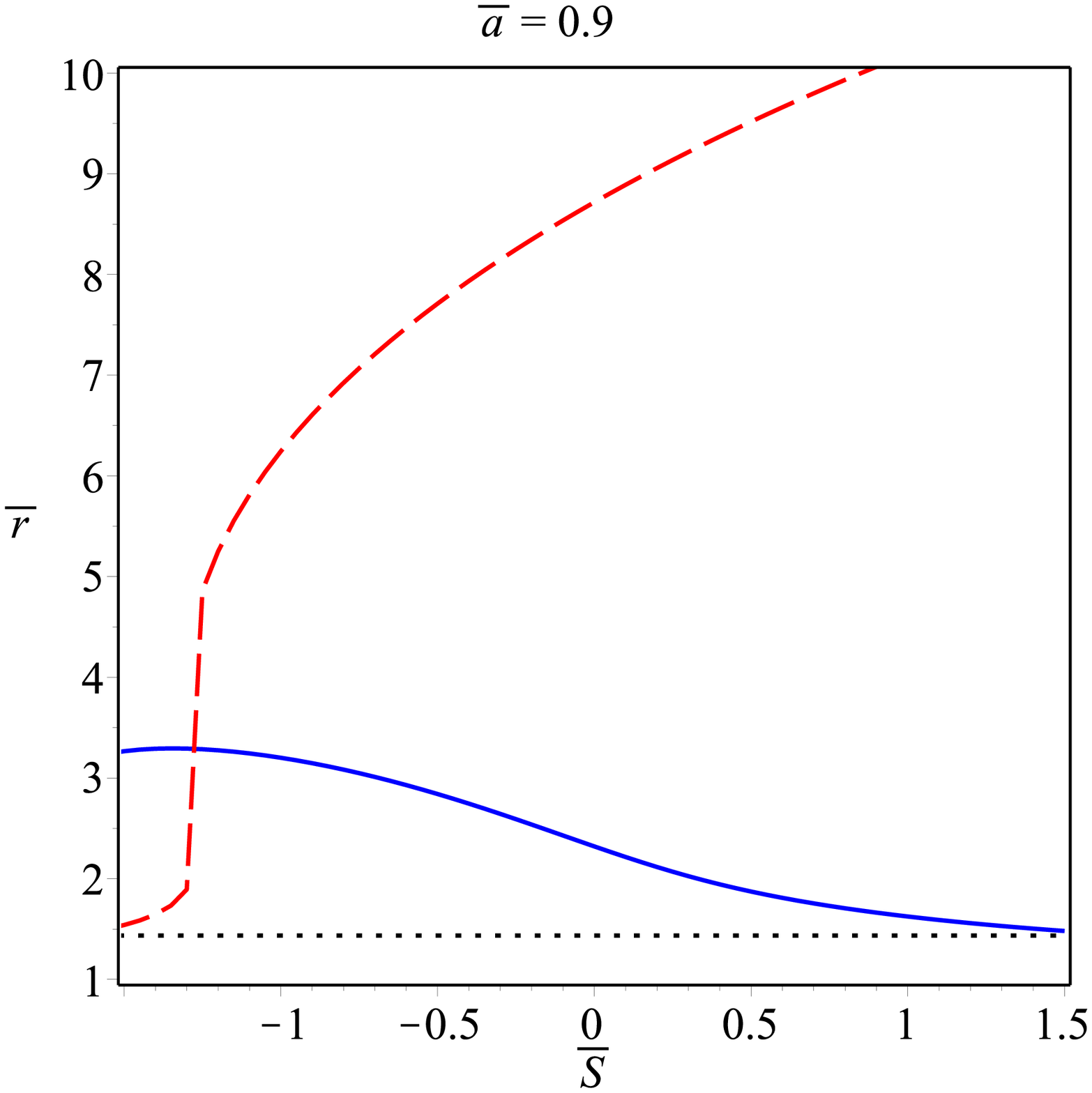}
\caption{Radius of the innermost radially stable circular orbit as a function of the spin $\bS$ for different values of the Kerr parameter $\ba$. The blue solid lines correspond to prograde orbits $\ba\bJ>0$ and the red dashed lines to retrograde orbits $\ba\bJ<0$. The event horizon is indicated by the black dotted line.}
\label{Fig:ISCO}
\end{figure*}

If in addition to the conditions \eqref{Cond_doubles} also $\frac{d^2\bar P_{\ba}}{d\br^2}<0$ holds, then the circular orbit is stable against radial perturbations. The radius of the innermost stable circular orbit (ISCO) with $\frac{d^2\bar P_{\ba}}{d\br^2}=0$ is of particular importance as it marks the transition from bound motion to infalling orbits. In figure \ref{Fig:ISCO} the radius of the innermost radially stable circular orbit is plotted as a function of the spin $\bS$ for fixed values of the Kerr rotation parameter $\ba$. Note that in general radially stable orbits may still be unstable against perturbations in the $\theta$-direction. Suzuki and Maeda \cite{SuzukiMaeda1998} have shown that radially stable circular orbits become unstable in the $\theta$ direction for large positive spin values $\bS\gtrsim 0.9$, but they only considered prograde motion ($\ba\bJ>0$). However, from figure \ref{Fig:ISCO} we infer that radially stable retrograde circular orbits with negative spins may come closer to the horizon than the corresponding prograde orbits (see also \cite{TodFelice1976}). Therefore, it would be interesting to analyze whether the same instabilities as reported in \cite{SuzukiMaeda1998} also appear for retrograde orbits with negative spins.

\subsection{General orbits}

\begin{figure*}
\includegraphics[width=0.32\textwidth]{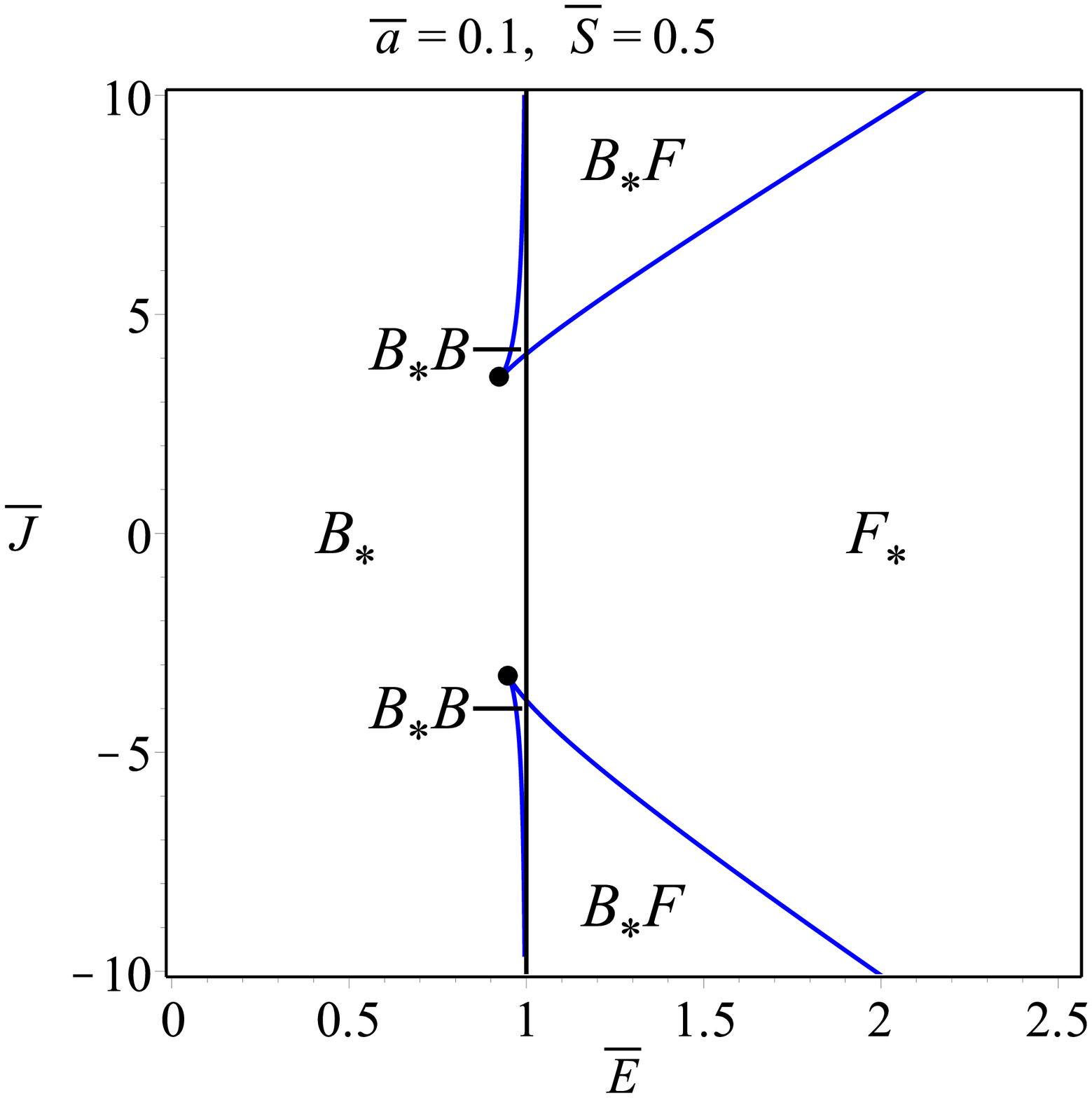}\quad
\includegraphics[width=0.32\textwidth]{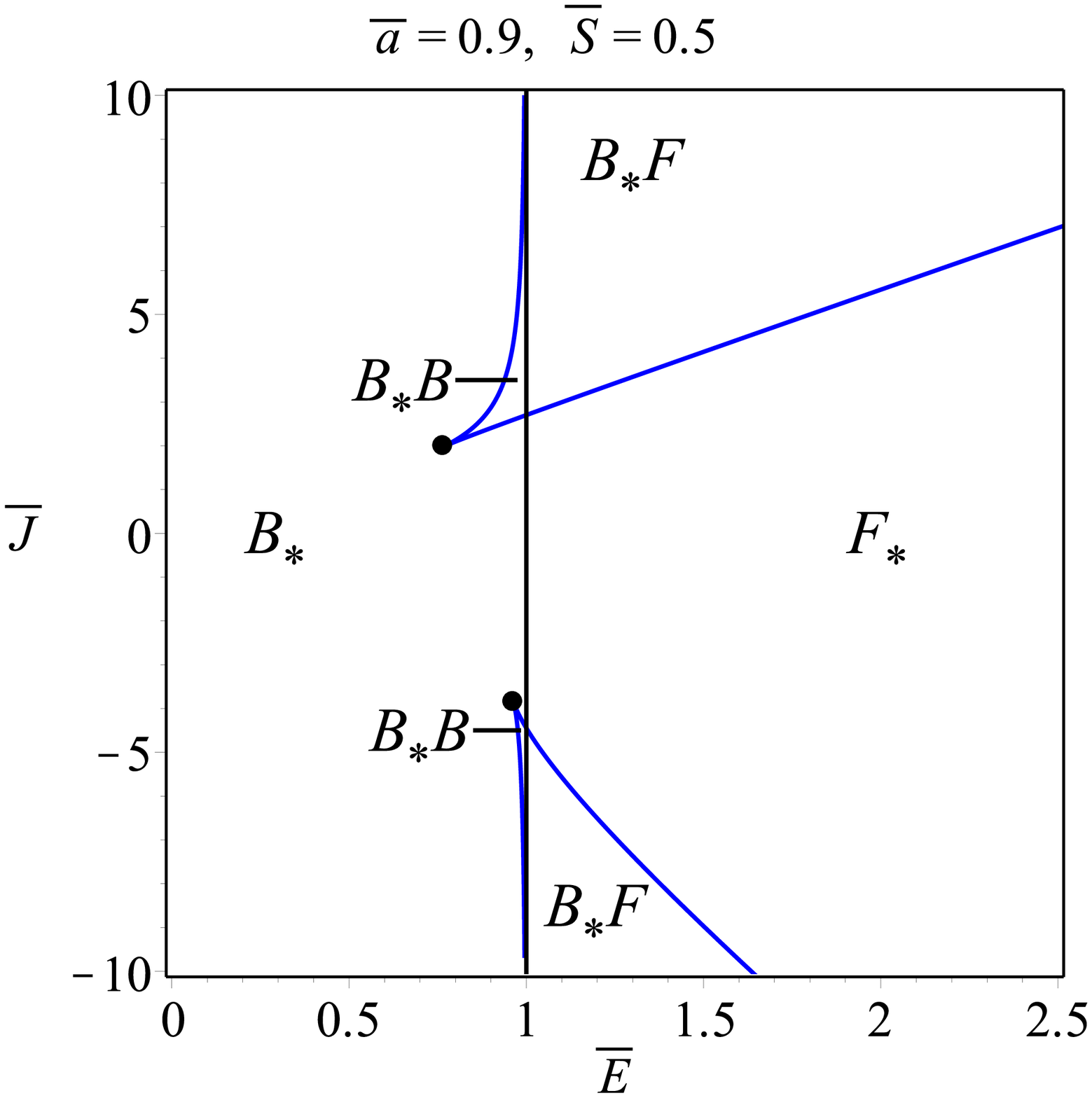}\quad
\includegraphics[width=0.32\textwidth]{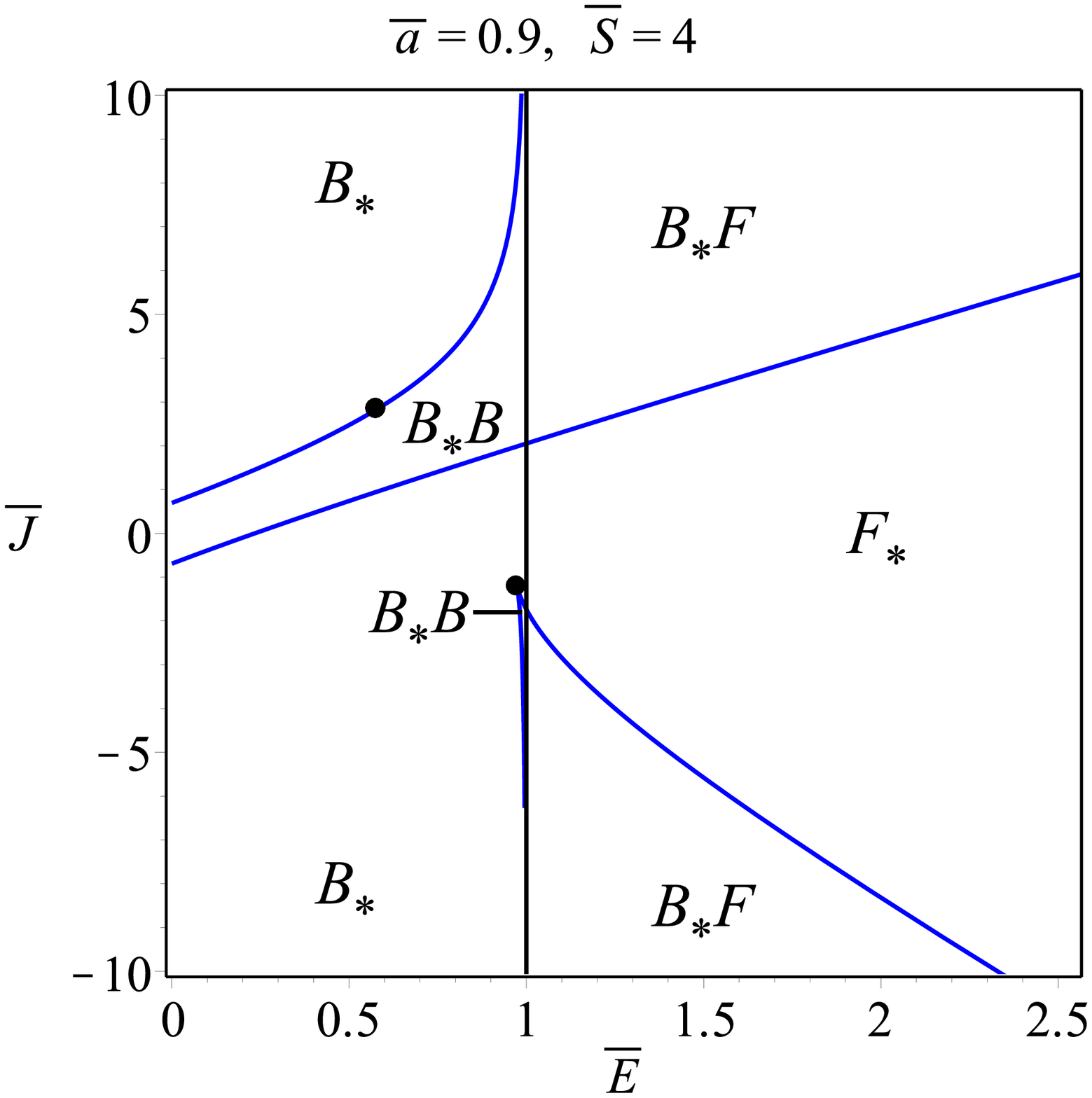}
\caption{Orbit types in parameter space for Kerr spacetime. The blue and black lines divide the plot in different regions of orbit types. Here, $B$ denotes a bound and $F$ a flyby orbit. A star indicates that the particle crosses the horizon. If more than one type of orbit is possible, the initial conditions determine the actual orbit. On the blue lines the orbits are circular and the dots indicate the innermost radially stable circular orbit. Only on the blue line approaching $\bE=1$ the circular orbits are radially stable. (Note that here only $\br\geq\br_+$ is considered; there are also combinations of $\bE$ and $\bJ$ where circular orbits between $\br=0$ and $\br=\br_-$ are possible.)}
\label{Fig:Kerr_EJ}
\end{figure*}

For given values of the parameters of the spacetime and the particle all possible types of motion are given by the regions where $\bar P_{\ba} \geq 0$, which can be directly inferred from the number of turning points $\bar P_{\ba}=0$ and the asymptotic behaviour of $\bar P_{\ba}$ at infinity. If we continuously vary the values of the parameters, the number of turning points changes at that set of parameters which correspond to double zeros, and which are given by \eqref{CircE} and \eqref{CircJ}. The asymptotic behaviour of $\bar P_{\ba}$ changes at $\bE^2=1$. Therefore, we may already infer all possible types of orbits from the analysis above. In Fig.~\ref{Fig:Kerr_EJ}, orbit types in parameter space are shown for fixed $\ba$ and $\bS$. Note that we only consider orbits which start at a radius $\br>\br_+=1+\sqrt{1-\ba^2}$ (motion outside the horizons).

\subsection{Schwarzschild spacetime}

\begin{figure*}
\includegraphics[width=0.32\textwidth]{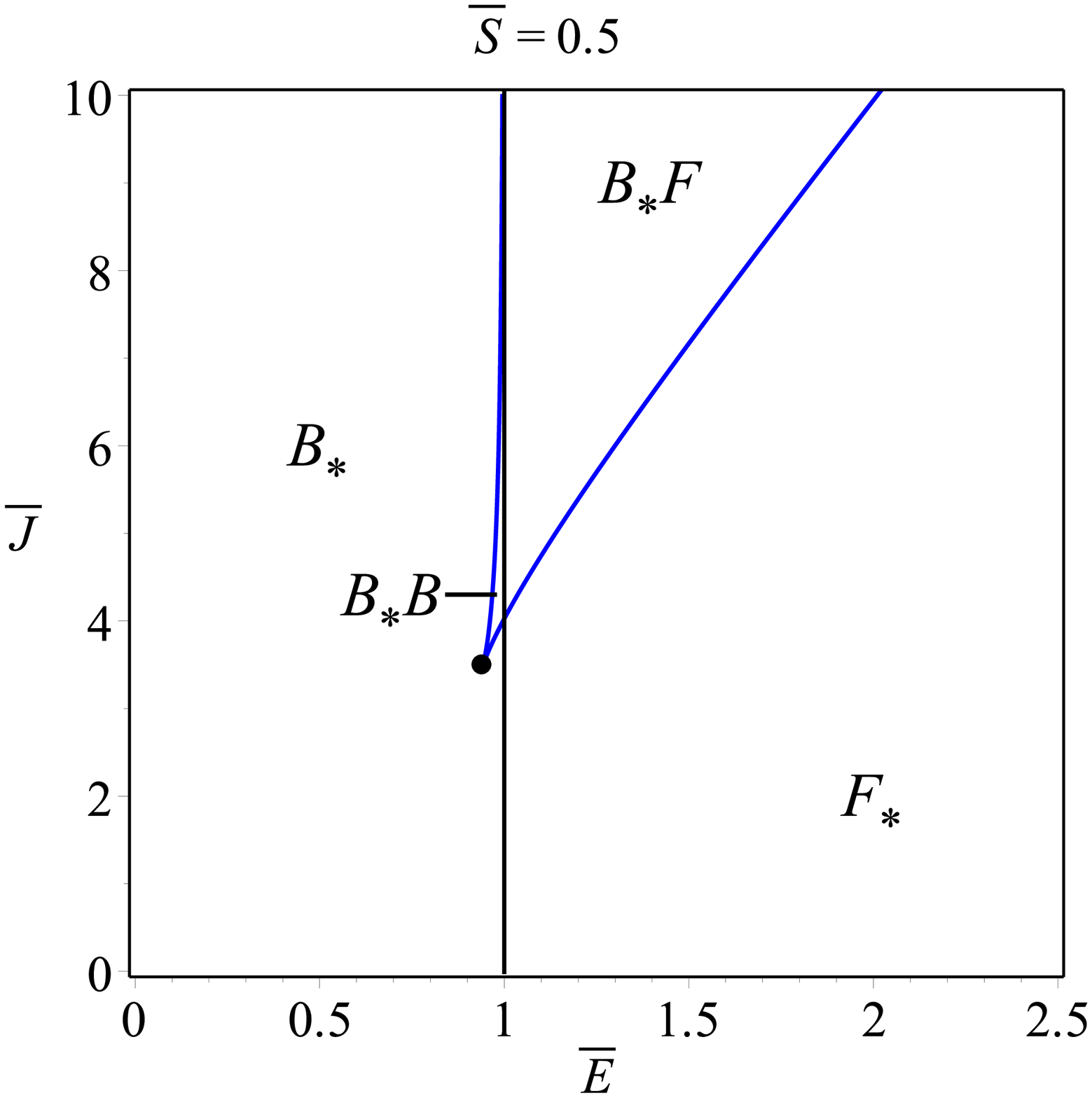}
\includegraphics[width=0.32\textwidth]{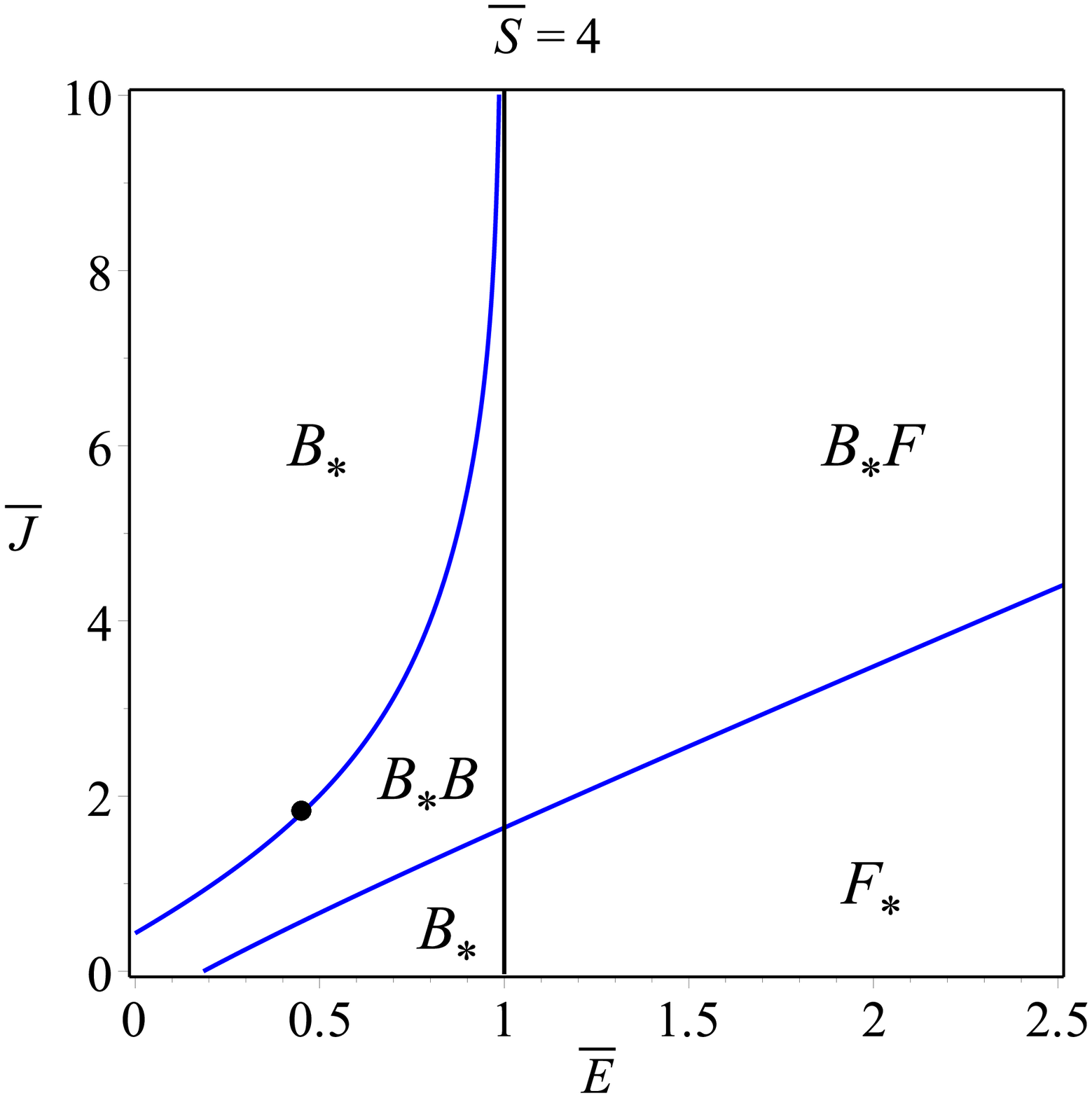}
\includegraphics[width=0.32\textwidth]{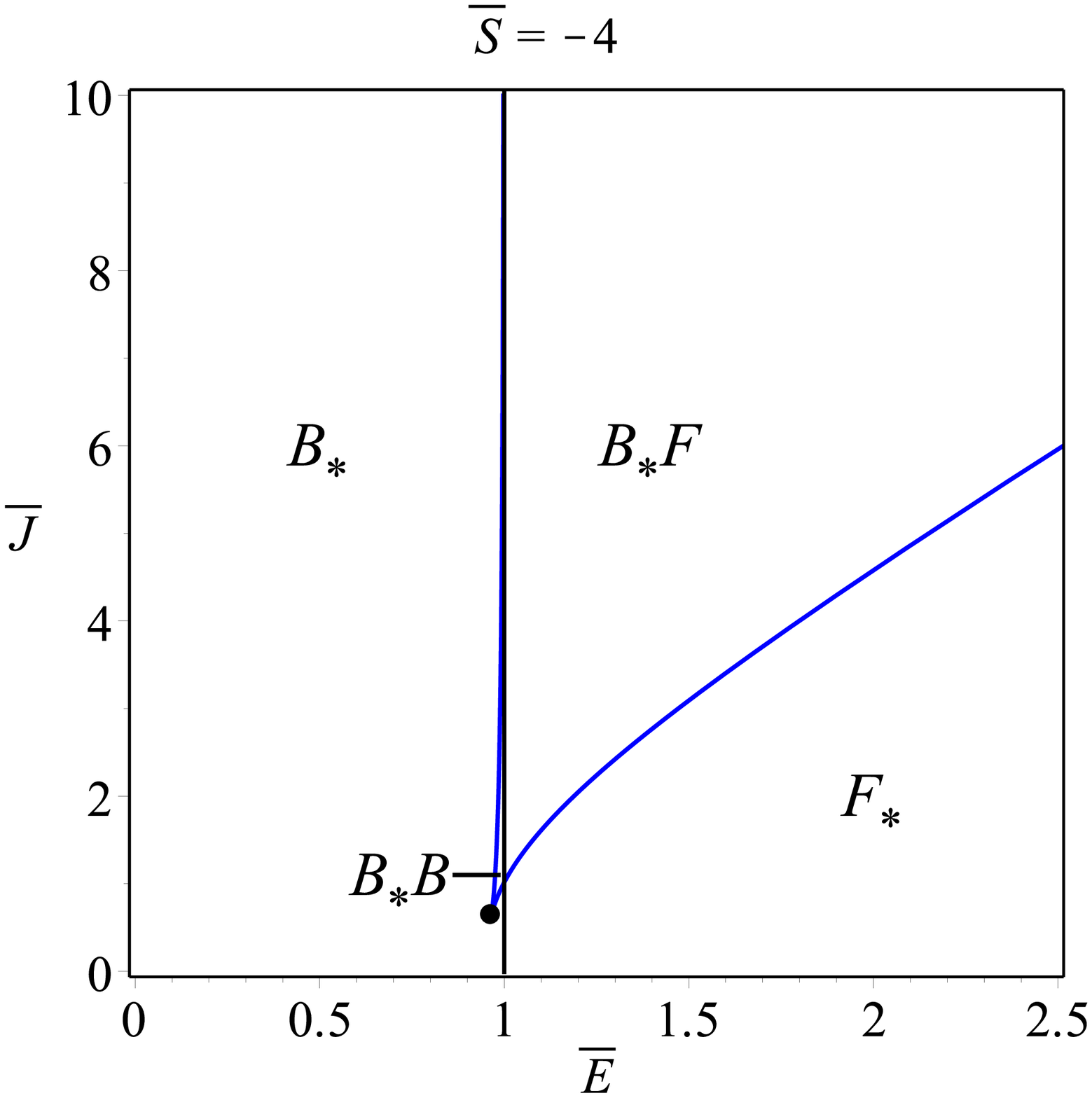}
\caption{Orbit types in parameter space for Schwarzschild spacetime. The blue and black lines divide the plot in four regions of different orbit types. Here, $B$ denotes a bound and $F$ a flyby orbit. A star indicates that the particle crosses the horizon. If more than one type of orbit is possible, the initial conditions determine the actual orbit. On the blue lines the orbits are circular and the dot indicates the innermost circular orbit which is stable in the radial direction. Only on the blue line from the dot approaching $\bE=1$ the circular orbits are radially stable.}
\label{Fig:Schwarzschild_EJ}
\end{figure*}

In Schwarzschild spacetime ($\ba=0$) the quotient of the velocity components simplifies to

\begin{align}
\frac{d\br}{d\phi} & = \frac{(\br^3+\bS^2)\sqrt{\bar P_{\ba=0}}}{\br^2(\br^3-2\bS^2)(\bJ-\bS\bE)}\,. \label{S_drdphinorm}
\end{align}
With $\ba=0$ the polynomial in \eqref{defPa} does not reduce its degree, but $\br=0$ is always a zero. The conditions \eqref{Cond_doubles} for $\ba=0$ are solved by
\begin{widetext}
\begin{align}
\bE_{1,2,3,4} & = \pm \frac{\sqrt{\br}\left[\bar V_{\ba=0}(\bar R_{\ba=0}\pm\br(\br-2)\sqrt{\bar U_{\ba=0}^2\br(4\br^6+13\br^2\bS^2-8\bS^4)})\right]^{\frac{1}{2}}}{\sqrt{2}\br^3 \bar V_{\ba=0}}\,,\\
\bJ_{1,2,3,4} & = \frac{\bS^6+\br^3(2\br-3)(\br\bE^2+\br-4)\bS^4-\br^6(\br^2+3\br\bE^2-7\br+9)\bS^2+\br^9(\br^2(\bE^2-1)-3\br\bE^2+4(\br-1))}{-\br^2\bE \bar U_{\ba=0}}\,,
\end{align}
\end{widetext}
Note that $(\bS,\bE)\to(-\bS,-\bE)$ does not change the equation of motion, so we may choose $\bE\geq0$. Also $(\bE,\bJ)\to(-\bE,-\bJ)$ only changes the sign, $\frac{d\br}{d\phi}\to-\frac{d\br}{d\phi}$. Therefore, this only reverses the direction but leaves the type of orbit unchanged, hence we choose $\bJ\geq0$. In Fig.~\ref{Fig:Schwarzschild_EJ} orbit types in parameter space are shown for fixed $\bS$.

\section{Periastron shift}\label{perihel_sec}

Let us consider the bound orbit of a spinning particle in the equatorial plane with turning points $\br_{\rm p}<\br_{\rm a}$. In this case $\bar P_{\ba}$ has at least four real zeros $\br_1<\br_2<\br_3<\br_4$ with $\br_3=\br_{\rm p}$ and $\br_4=\br_{\rm a}$. The periastron shift $\Delta \omega=2\pi(K-1)$ is then given by the difference of the periodicity of the radial motion $\br(\phi)$ and $2\pi$, i.e.
\begin{align}
K & = \frac{1}{\pi} \int_{\br_{\rm p}}^{\br_{\rm a}} \frac{f }{\sqrt{P_{\ba}}}\,d\br , \label{DefPerihel}
\end{align}
where $f=\frac{\br \bar Q}{\bar{\Delta}(\br^3 + \bS^2)}$. 

Let us write $P_{\ba}=(\bE^2-1)\prod_{i=1}^8(\br-\br_i)$ where $\br_1<\ldots<\br_4 \in \mathds{R}$ with $\br_3=\br_{\rm p}$, $\br_4=\br_{\rm a}$, and $\br_5,\ldots,\br_8 \in \mathds{C}$. If we introduce a new variable $z$ by $\br=\frac{(\br_{\rm a}-\br_1)(\br_{\rm p}-\br_1)}{z(\br_{\rm a}-\br_{\rm p})+\br_{\rm p}-\br_1} + \br_1$ the expression \eqref{DefPerihel} transforms to
\begin{align}
K & = \frac{(\br_{\rm a}-\br_1)^2}{\pi \sqrt{(1-\bE^2)\prod_{i=5}^8(\br_{\rm a}-\br_i)(\br_{\rm p}-\br_1)(\br_{\rm a}-\br_2)}} \nonumber\\
& \quad \times \int_0^1 \frac{\sum_{i=0}^2 C_iz^i + \sum_{j=1}^5 \frac{B_{j}}{1-b_{j}z}}{\sqrt{z(1-z)\prod_{i=1}^5(1-l_iz)}} \,dz\,, \label{perihelintegrals}
\end{align}
where the constants $C_i$, $B_{j}$ are the coefficients of a partial fraction expansion of $f(\br(z))/(\br(z)-\br_1)^2$, which are given in appendix \ref{app:FD}, together with the characteristics $b_{j}$. The parameter $\vec{l}=(l_1,\ldots,l_5)$ is defined as
\begin{align}
l_1 & =\frac{(\br_{\rm a}-\br_{\rm p})(\br_2-\br_1)}{(\br_{\rm p}-\br_1)(\br_{\rm a}-\br_2)}\,, \label{defl1} \\
l_i & = \frac{(\br_{\rm a}-\br_{\rm p})(\br_{i+3}-\br_1)}{(\br_{\rm p}-\br_1)(\br_{\rm a}-\br_{i+3})}\,, \quad i=2,\ldots,5\,. \label{defli}
\end{align}
The expression \eqref{perihelintegrals} can now be rewritten in terms of Lauricella's hypergeometric $F_D$ function, which is given in terms of a power series and can be calculated quite easily, see appendix \ref{app:FD} for details,
\begin{align}
K & = \frac{(\br_{\rm a}-\br_1)^2}{\sqrt{(1-\bE^2)\prod_{i=5}^8(\br_{\rm a}-\br_i)(\br_{\rm p}-\br_1)(\br_{\rm a}-\br_2)}} \nonumber\\
& \times \bigg[ \frac{3C_2}{8} F_D\left(\frac{5}{2},\vec{\beta},3,\vec{l}\right) + \frac{C_1}{2} F_D\left(\frac{3}{2},\vec{\beta},2,\vec{l}\right) \nonumber\\
& + C_0 F_D\left(\frac{1}{2},\vec{\beta},1,\vec{l}\right) + \sum_{j=1}^5 B_{j} F_D\left(\frac{1}{2},\vec{\beta}^*,1,\vec{l}^*_{b_j}\right) \bigg]\,, \label{exactperihel}
\end{align}
where $\vec{\beta}=(\beta_1,\ldots,\beta_5)$ with $\beta_j=1/2$ for all $j$, $\vec{\beta}^*=(\beta_1,\ldots,\beta_5,1)$, and $\vec{l}^*_{b_j}=(l_1,\ldots,l_5,b_j)$.

For $\bS=0$ expression \eqref{exactperihel} reduces to the known exact expressions for the periastron shift of equatorial motion of a spinless test body in Kerr spacetime. For $\bS=0$ we see that five zeros of $\bar P_{\ba}$ coincide, $\br_1=0=\br_i$ for $5\leq i \leq8$, and, therefore, the expressions \eqref{defli} vanish. In addition, $C_i=0$, $i=0,1,2$, and $B_3=B_4=B_5=0$ for $\bS=0$. Then $F_D\left(\frac{1}{2},\vec{\beta}^*,1,\vec{l}^*_{b_j}\right)=2\Pi(b_{j},\sqrt{l_1})/\pi$, $j=1,2$, where $\Pi$ is the complete elliptic integral of the third kind, and we find
\begin{align}
K(\bS=0) & = \sum_{j=1}^2 \frac{2 B_j \Pi(b_{j},\sqrt{l_1})}{\pi \sqrt{(1-\bE^2) \br_{\rm p} (\br_{\rm a}-\br_2)}}\,.
\end{align}
If in addition $\ba=0$ we find $B_2=0$, $B_1=\bJ$, and $b_{1}=0$ which gives
\begin{align}
K(\bS=0,\ba=0) & = \frac{2\bJ \mathcal{K}(\sqrt{l_1})}{\pi\sqrt{(1-\bE^2) \br_{\rm p} (\br_{\rm a}-\br_2)}}\,,
\end{align}
where $\mathcal{K}(\sqrt{l_1})=\Pi(0,\sqrt{l_1})$ is the complete elliptic integral of the first kind. 

\begin{figure}
\centering
\includegraphics[width=0.46\textwidth]{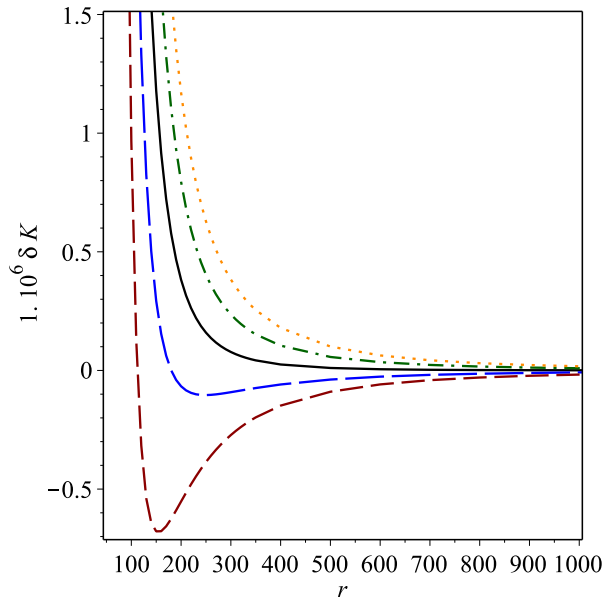}
\caption{Comparison of post-Newtonian and exact expression for the periastron shift of quasi-circular orbits. Here the difference $\delta K$ (multiplied by $10^6$), i.e.~the exact expression \eqref{qc_exactperihel} minus the post-Newtonian expression \eqref{qc-PN-perihel},  is shown as a function of the radius $r$ of the quasi-circular orbit ($r$ in units of $M$). We chose the spin of the Kerr black hole as $\ba=0.9$. The black solid line corresponds to a spin zero test particle, the blue long dashed line to $\bS=0.1$, the red short dashed line to $\bS=0.2$, the green dash dotted line to $\bS=-0.1$, and the orange dotted line to $\bS=-0.2$.}
\label{Fig:quasi-circ}
\end{figure}

For spinning black hole binaries in quasi-circular orbits the post-Newtonian expansion of the periastron precession was determined in \cite{LeTiecetal2013}, see also \cite{Tessmer:etal:2013}. In \cite{LeTiecetal2013} the test particle limit in the pole-dipole-quadrupole approximation (their equation (24)) was considered, which reads for the periastron shift
\begin{align}
K & = \bigg[ 1 - \frac{6}{\br} + \frac{8\ba+6\bS}{\br^{\frac32}} - \frac{3\ba^2+6\ba\bS}{\br^2}\nonumber\\
& \quad - \frac{18\bS}{\br^\frac52} + \frac{30\ba\bS}{\br^3} - \frac{12\ba^2\bS}{\br^\frac72} + \mathcal{O}(\bS^2)\bigg]^{-\frac12}\,. \label{qc-PN-perihel}
\end{align}
Our expression \eqref{exactperihel} for $\br_{\rm a}=\br_{\rm p}$ reduces to 
\begin{align}
K(\br_{\rm a}=\br_{\rm p}) & = C_0 + \sum_{j=1}^5 B_j\,. \label{qc_exactperihel}
\end{align}
A comparison of these expressions is visualized in figure \ref{Fig:quasi-circ}. In the region of large $r$, where the post-Newtonian approximation is valid, the two expressions coincide very well. For smaller values of $r$, say around $r=500M$ in fig.~\ref{Fig:quasi-circ}, the approximate formula \eqref{qc-PN-perihel} works still quite well for vanishing spin but shows already quite significant deviations for larger values of $\bS$. Here the quadratic effects of the spin, which are neglected in \eqref{qc-PN-perihel}, apparently become already important, at least in combination with the post-Newtonian approximation.

\begin{figure}
\centering
\includegraphics[width=0.46\textwidth]{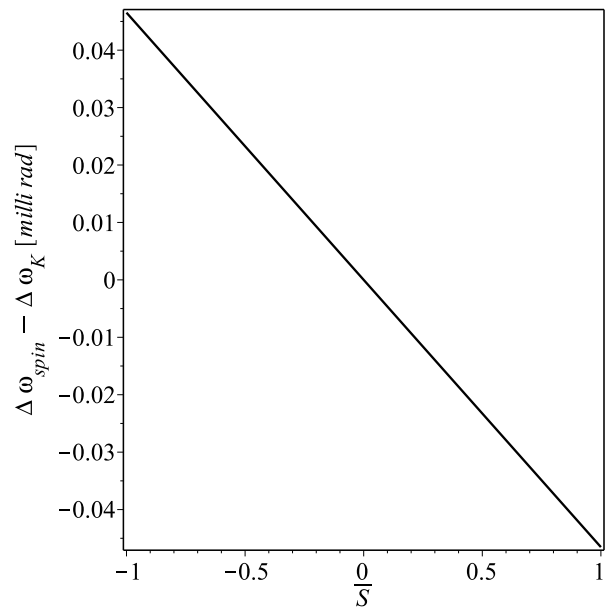}
\caption{The relativistic periastron shift of a spinning SO-2 in the equatorial plane of a Kerr black hole with rotation $\ba=0.95$. Here the difference between the periastron shift of a non-spinning SO-2 in a prograde orbit, denoted by $\Delta\omega_{\rm K}$, and the periastron shift including the spin, denoted by $\Delta\omega_{\rm spin}$, is shown in $10^{-3}$ rad as a function of the dimensionless  spin $\bS$.}
\label{Fig:S2spin}
\end{figure}

Let us apply the exact expression \eqref{exactperihel} to a stellar orbit around Sagittarius A*, the massive black hole at the center of our galaxy, to get an impression about the magnitude of the spin effects. We consider here SO-2 which has well known orbital parameters and a very short orbital period. From \cite{Gillessenetal2009} we take the eccentricity $e_{\rm S2}=0.88$, the semi major axis $d_{\rm S2}=0.123 \, \rm mas$, the mass of the black hole as $M_{\rm BH}=4.31 \times 10^6 \, M_{\rm Sun}$ and its distance as $R_{\rm BH}=8.33 \,\rm kpc$. With $M_{\rm Sun}\approx1476.9\,\rm m$ we derive from this the normalized peri- and apastron of SO-2 as
\begin{align}
\br_{\rm a} & \approx 45.27 \times 10^{3}\,,\\
\br_{\rm p} & \approx 2.89 \times 10^{3}\,.
\end{align}
If we assume that SO-2 moves in the equatorial plane of a Kerr black hole we may use \eqref{exactperihel} to derive the relativistic periastron precession including the effect of a possible spin of SO-2. The usual post-Newtonian formula for the relativistic precession per orbital period is 
\begin{align}
\Delta \omega_{\rm PN} & = \frac{6\pi M_{\rm BH}}{d_{\rm S2}(1-e_{\rm S2}^2)} \approx 3.47 \times 10^{-3} \, \rm rad \approx 11.9'\,.
\end{align}
This value is in good agreement with \eqref{exactperihel} for $\ba=0$ and $\bS=0$, $\Delta\omega(\ba=0,\bS=0)-\Delta\omega_{\rm PN} \approx 3\times 10^{-6} \, \rm rad$ (remember $\Delta\omega=2\pi(K-1)$). If we assume a black hole spin of $\ba=0.95$ and a non-spinning SO-2 in a prograde orbit we get a correction of $\Delta\omega(\ba=0.95,\bS=0)-\Delta\omega_{\rm PN} \approx -5.94 \times 10^{-5}\,\rm rad$. In figure \ref{Fig:S2spin} we see the effect of a non vanishing spin of SO-2 as compared to the case of $\bS=0$. For large SO-2 spins the correction is nearly of the same order as the correction due to the black hole spin, approximately $\mp 4.65 \times 10^{-5}$ radians per orbital period for $\bS=\pm 1$. For a comparison of Newtonian and post-Newtonian contributions see also \cite{Rubilar:Eckart:2001}.

\section{Conclusions}\label{conclusions_sec}

In this paper we derived an explicit velocity formula for particles in the equatorial plane of a Kerr black hole with spin aligned or anti-aligned with the rotation of the black hole. We classified the radial motion of such particles outside the horizons and also plotted the location of the innermost radially stable circular orbit. From figure \ref{Fig:ISCO} it can be inferred that not only prograde orbits but for large negative spins also retrograde orbits may come very close to the outer black hole horizon. It would be interesting to analyze whether such retrograde orbits also become unstable in the $\theta$-direction as shown for prograde orbits in \cite{SuzukiMaeda1998}. 

We then used the explicit velocity formula to derive an exact expression for the periastron shift of a spinning particle. A comparison with a post-Newtonian expression for quasi-circular orbits up to first order in the spin given in \cite{LeTiecetal2013} showed that the quadratic spin contributions should be included before adding even higher order post-Newtonian terms to this expression. In order to get an idea about the magnitude of spin corrections to the periastron shift we considered as an example the orbit of SO-2 around the galactic center black hole. Assuming prograde equatorial motion and an (anti-)aligned spin of SO-2 we found that the corrections due to a  spinning SO-2 may become nearly as large as the corrections due to the spin of the black hole. Therefore, this effect may become relevant for tests of General Relativity in the vicinity of the central black hole using stellar orbits.

\begin{acknowledgments}
This work was supported by the Deutsche Forschungsgemeinschaft (DFG) through the grants LA-905/8-1/2 (D.P.), DI-527/6-1 (I.S.), and the research training group 'Models of Gravity' (E.H.). We would like to thank V.\ Perlick and B.\ Mashhoon for fruitful discussions.
\end{acknowledgments}

\appendix

\section{Conventions \& Symbols}\label{dimension_acronyms_app}

\begin{table}
\caption{\label{tab_dimensions}Dimensions of the quantities.}
\begin{ruledtabular}
\begin{tabular}{cp{0.3\textwidth}}
Dimension (SI)&Symbol\\
\hline
&\\
\hline
\multicolumn{2}{l}{{Geometrical quantities}}\\
\hline
1 & $g_{ab}$, $\sqrt{-g}$, $\delta^a_b$, $\varepsilon^{a b c d}$, $\theta$, $\phi$, $d\theta$, $d\phi$\\
m& $s$, $Y^a$, $t$, $r$, $dt$, $dr$, $\rho$, $\Delta$, $M$, $a$, $r_{\pm}$, $M_{\rm BH}$, $M_{\rm Sun}$, $R_{\rm BH}$ \\
m$^{-2}$& $R_{abcd}$ \\
$[E_\xi]$ kg$^{-1}$ & ${\xi^{a}}$ \\
&\\
\hline
\multicolumn{2}{l}{{Matter quantities}}\\
\hline
1& $u^a$, $K$, $e_{\rm S2}$, $d_{\rm S2}$\\
kg& $m$, $\underline{m}$, $p^a$, $E$ \\
kg\,m& $S^{ab}$, $S^{a}$, $S$, $J$ \\
rad & $\Delta\omega$, $\Delta\omega_{\rm PN}$\\
&\\
\hline
\multicolumn{2}{l}{{Auxiliary quantities}}\\
\hline
1 & $\hat{p}^a$, $\bar P_{\ba}$, $\bar Q$, $\bar V_{\ba}$, $\bar R_{\ba}$, $\bar U_{\ba}$, $\br_{1,\ldots,8}$, $z$, $C_{0,\ldots,2}$, $B_{1,\ldots,5}$, $b_{1,\ldots,5}$, $l_{1,\ldots,5}$, $\vec{\beta}$, $\vec{\beta}^*$, $\vec{l}$, $\vec{l}^*_{b_j}$ \\
kg & $\mu$ \\
kg$^{-2}$ & $A_1$ \\
&\\
\hline
\multicolumn{2}{l}{{Operators \& functions}}\\
\hline
1 & $F_D$, $\Gamma$, $\Pi$, $\mathcal{K}$  \\ 
m$^{-1}$& $\partial_a$, $\nabla_i$, $\frac{D}{ds} = $``$\dot{\phantom{a}}$'' \\
&\\
\end{tabular}
\end{ruledtabular}
\end{table}

\begin{table}
\caption{\label{tab_symbols}Directory of symbols.}
\begin{ruledtabular}
\begin{tabular}{lp{0.3\textwidth}}
Symbol & Explanation\\
\hline
&\\
\hline
\multicolumn{2}{l}{{Geometrical quantities}}\\
\hline
$g_{a b}$ & Metric\\
$\sqrt{-g}$ & Determinant of the metric \\
$\delta^a_b$ & Kronecker symbol \\
${\xi^a}$ & Killing vector\\
$t$, $r$, $\theta$, $\phi$  & Coordinates\\
$s$ & Proper time \\
$Y^a$ & Worldine\\ 
$R_{a b c d}$& Curvature \\
$M$, $a$ & Kerr (mass, parameter) \\ 
$r_+$, $r_-$ & (outer, inner) horizon \\
$M_{\rm BH}$, $M_{\rm Sun}$ & Mass (black hole, Sun)\\
$R_{\rm BH}$ & Distance to black hole\\
&\\
\hline
\multicolumn{2}{l}{{Matter quantities}}\\
\hline
$u^a$ & Velocity \\
$m$, $\underline{m}$ & Mass (Frenkel, Tulczyjew)\\
$p^a$ & Generalized momentum \\
$E_\xi$ & General conserved quantity\\
$S^{ab}$, $S^a$, $S$ & Spin (tensor, vector, length) \\
$E$, $J$ & Energy, angular momentum \\
$\br_{\rm a}$, $\br_{\rm p}$ & apastron, periastron\\
$K$, $\Delta\omega$, $\Delta\omega_{\rm PN}$ & periastron advance (dimensionless, in rad, in PN-approximation) \\
$e_{\rm S2}$, $d_{\rm S2}$ & SO-2 eccentricity, semi-major axis\\
&\\
\hline
\multicolumn{2}{l}{{Operators \& functions}}\\
\hline
$\varepsilon^{abcd}$ & Permutation symbol\\
$\partial_i$, $\nabla_i$, $\frac{D}{ds} = $``$\dot{\phantom{a}}$''& (Partial, covariant, total) derivative \\ 
``$\bar{\phantom{A}}$'' & Dimensionless quantity \\
$F_D$ & Lauricella function \\
$\Gamma$ & Gamma function \\
$\mathcal{K}$, $\Pi$ & Complete elliptic integrals (first \& third kind)\\
&\\
\end{tabular}
\end{ruledtabular}
\end{table}

The dimensions of the different quantities appearing throughout the work are displayed in table \ref{tab_dimensions}. We set $c=1$, the dimension of the gravitational constant then becomes $[G]=\rm m/kg$. Table \ref{tab_symbols} contains a list with the most important symbols used throughout the text. Latin indices denote 4-dimensional indices and run from $a = 0, \dots, 3$, the signature is (+,--,--,--). 

\section{Lauricella's $F_D$ function}\label{app:FD}

The four functions $F_A$, $F_B$, $F_C$, and $F_D$ of Lauricella are hypergeometric functions of multiple variables generalizing the hypergeometric functions of Gauss and Appell. They were introduced in 1893 \cite{Lauricella1893} and given as a hypergeometric series
\begin{align}
F_D(\alpha,\vec{\beta},\gamma,\vec{x}) & = \sum_{\vec{\iota}=0}^{\infty} \frac{(\alpha)_{|\vec{\iota}|}\,(\vec{\beta})_{\vec{\iota}}}{(\gamma)_{|\vec{\iota}|}\, \vec{\iota}!} \,\, \vec{x}^{\,\vec{\iota}}\,,
\end{align}
where $\vec{\iota}$ is a multi-index, $|x_{\iota_n}|<1$ for all $n$, and $(\cdot)$ is the Pochhammer symbol. Here $|\vec{\iota}|=\sum_n \iota_n$, $\vec{\iota}!=\prod_n \iota_n!$, and $(\vec{\beta})_{\vec{\iota}} = \prod_n (\beta_n)_{\iota_n}$. The function $F_D$ can be extended to other values of $\vec{x}$ by analytic continuation. It can also be rewritten as a simple series which is much more convenient for computations \cite{vanLaarhovenetal1988},
\begin{align}
F_D(\alpha,\vec{\beta},\gamma,\vec{x}) & = 1+ \sum_{m=1}^{\infty} \frac{(\alpha)_m}{(\gamma)_m} \varLambda_m\,,
\end{align}
where
\begin{align}
\varLambda_m & = \sum_{|\vec{\iota}|=m} \frac{(\vec{\beta})_{\vec{\iota}}}{\vec{\iota}!}\, \vec{x}^{\,\vec{\iota}}\nonumber\\
& = \sum_{\{\vec{m} \in \mathds{N}^n|\sum_j m_j=m\}} \, \prod_{j=1}^n \frac{(\beta_j)_{m_j}}{m_j!} \, x_j^{m_j}\,.
\end{align}

In this paper the $F_D$ function is used because it can be represented in an integral form
\begin{align}
& F_D(\alpha,\vec{\beta},\gamma,\vec{x}) = \frac{\Gamma(\gamma)}{\Gamma(\alpha)\Gamma(\gamma-\alpha)}\nonumber\\
& \quad \times \int_0^1 t^{\alpha-1} (1-t)^{\gamma-\alpha-1} \prod_n (1-x_nt)^{-\beta_n} dt 
\end{align}
for ${\textrm Re}(\gamma)>{\textrm Re}(\alpha)>0$, where $\Gamma$ denotes the gamma function. It is a generalization of the Jacobian elliptic integrals, e.g. $\pi F_D(1/2,1/2,1,k^2)=2\mathcal{K}(k)$, where $\mathcal{K}$ is the complete elliptic integral of the first kind.

The constants appearing in front of $F_D$ in \eqref{exactperihel} are due to a partial fraction decomposition of $f(\br(z))/(\br(z)-\br_1)^2$ and given by
\begin{align}
B_{1,2} & = \frac{\ba(\br_{\rm a}-\br_1)}{2\bar \Delta(\br_1)^3(\ba^2+\br_{\rm a}(\br_1-2)+(\br_{\rm a}-\br_1)r_{\pm})(1-r_{\pm})} \nonumber \\
& \quad \times \big[ (\bJ-\bS\bE)\ba^7 -2\bE (r_{\pm}+4\br_1-2)\ba^6\nonumber\\
& \quad + [(2\bJ-\bE\bS-4(\bJ-\bE\bS)\br_1)r_{\pm}-6(\bJ-\bE\bS)\br_1^2\nonumber\\
& \quad  +4(2\bJ-\bE\bS)\br_1-4\bJ]\ba^5 + 2\bS\ba^2(\bJ-\ba\bE)(2(\br_1-1)^3 \nonumber\\
& \quad +2r_{\pm}(\br_1-1)^2+r_{\pm}\br_1^2+2\br_1-2)\nonumber\\
& \quad +\br_1^3(\bJ-\bE\bS)(4+\br_1)\ba^3-2\br_1^3(\br_1+8)\bE r_{\pm}\ba^2\nonumber\\
& \quad +[(12\bE\br_1^2-\bJ\bS)r_{\pm}+8\bE\br_1^3-4\bJ\bS\br_1+4\bJ\bS]\ba^4 \nonumber\\
& \quad -\br_1^4(2\bJ-3\bE\bS)r_{\pm}\ba +\br_1^4(8\bE-\bJ\bS)r_{\pm} \big]
\end{align}
\begin{align}
B_{j} & = \big\{{\bS^2b_j(\ba\bE+\bS\bE-\bJ)(\br_{\rm a}-\br_1)(\br_{\rm p}-\br_1)}/(\bS^2+\br_1^3)^3\nonumber\\
& \quad \times (\br_{\rm a}-\br_{\rm p})[(\br_{\rm a}-\br_{\rm p}+b_j(\br_{\rm p}-\br_1))^2\bS^2\nonumber\\
& \quad+\br_1(\br_1(\br_{\rm p}-\br_{\rm a})+\br_{\rm a}b_j(\br_{\rm p}-\br_1))^2]\big\}\nonumber\\
& \quad \times \big[(\br_{\rm a}-\br_{\rm p})^2\br_1^2(10\bS^4-16\bS^2\br_1^3+\br_1^6)\nonumber\\
& \quad+(\br_{\rm a}-\br_{\rm p})(\br_{\rm p}-\br_1)\br_1(5(\br_{\rm a}+3\br_1)\bS^4\nonumber\\
& \quad-4\br_1^3(5\br_{\rm a}+3\br_1)\bS^2+2\br_{\rm a}\br_1^6)b_j\nonumber\\
& \quad+(\br_{\rm p}-\br_1)^2((\br_{\rm a}^2+3\br_{\rm a}\br_1+6\br_1^2)\bS^4\nonumber\\
& \quad-\br_1^3(7\br_{\rm a}+6\br_{\rm a}\br_1+3\br_1^2)\bS^2+\br_{\rm a}^2\br_1^6)b_j^2 \big], 
\end{align}
where $j=3,4,5$ and $r_{\pm}=1\pm\sqrt{1-\ba^2}$ are the horizons, and the characteristics $b_{j}$ are solutions of the equations
\begin{align}
0 & = 2(\br_{\rm a}-\br_{\rm p})(\br_{\rm p}-\br_1)(\ba^2-\br_{\rm a}-\br_1+\br_1\br_{\rm a})b_{j}\nonumber\\
& \quad +(\br_{\rm p}-\br_1)^2\bar \Delta(\br_{\rm a})b_{j}^2 +(\br_{\rm a}-\br_{\rm p})^2\bar \Delta(\br_1)\,,
\end{align}
for $j=1,2$ and 
\begin{align}
0 & = (\br_{\rm a}-\br_{\rm p}+b_{j}(\br_{\rm p}-\br_1))^3S^2\nonumber\\
& \quad +(\br_1(\br_{\rm a}-\br_{\rm p})+b_{j}\br_{\rm a}(\br_{\rm p}-\br_1))^3\,,
\end{align}
for $j=3,4,5$. Furthermore we have:
\begin{align}
C_2 & = \frac{(\br_{\rm a}-\br_{\rm p})\br_1}{\bar \Delta(\br_1)(\bS^2+\br_1^3)(\br_{\rm a}-\br_1)^2(\br_{\rm p}-\br_1)^2} \big[3\bE\bS^2\br_1\ba^3 \nonumber \\
&+(\bE\bS^2(1+3\br_1)-3\bS\bJ\br_1+\bE\br_1^3)\bS\ba^2 \nonumber\\
& \quad -(\bS^3\bJ+\bS^2\bE\br_1^2(4-3\br_1)+\bS\bJ\br_1^3-2\bE\br_1^5)\ba \nonumber \\
&+\br_1^2(\br_1-2)(\br_1^3-2\bS^2)(\bJ-\bE\bS)\big]\,, 
\end{align}

\begin{widetext}
\begin{align}
C_0 & = \frac{1}{\bar \Delta(\br_1)^3(\bS^2+\br_1^3)^3(\br_{\rm a}-\br_1)^2} \big[3\bE\bS^2(\br_{\rm a}\bS^4+(15\br_{\rm a}\br_1-7\br_{\rm a}^2-6\br_1^2)\bS^2+\br_1^6(\br_{\rm a}^2+3\br_1^2-3\br_{\rm a}\br_1))\ba^7\nonumber\\
& \quad + (\br_{\rm a}\bE\bS(\bS^6+(45\br_1+3)\br_1^3\bS^4+3(1-3\br_1)\br_1^6\bS^2+\br_1^9)-9\br_{\rm a}\bS^2\bJ\br_1^4(5\bS^2-\br_1^3)\nonumber\\
& \qquad -\bS^2(\bJ-\bE\bS)(3\br_{\rm a}^2\bS^4-3\br_1^3\bS^2(7\br_{\rm a}^2+6\br_1^2)+3\br_1^6(\br_{\rm a}^2+3\br_1^2)))\ba^6\nonumber\\
& \quad +(\br_1\bE((2\br_1^2+9\br_1\br_{\rm a}^2-12\br_{\rm a}^2-6\br_{\rm a}\br_1)\bS^6-3\br_1^3(18\br_1^3-45\br_{\rm a}\br_1^2-38\br_1^2+96\br_{\rm a}\br_1+21\br_1\br_{\rm a}^2-48\br_{\rm a}^2)\bS^4\nonumber\\
& \qquad +3\br_1^7(3\br_{\rm a}^2+12\br_{\rm a}-16\br_1-9\br_{\rm a}\br_1+9\br_1^2)\bS^2+2\br_1^9(3\br_{\rm a}^2+\br_1^2-3\br_{\rm a}\br_1))-\bJ\bS\br_{\rm a}(\bS^2+\br_1^3)^3)\ba^5\nonumber\\
& \quad + \{ \bE(8\br_{\rm a}\br_1^3+\br_1^3+3\br_{\rm a}^2\br_1^2-6\br_{\rm a}\br_1-15\br_1\br_{\rm a}^2+2\br_{\rm a}^2-3\br_1^4)\bS^7-\bJ\br_1(-6\br_{\rm a}\br_1+2\br_1^2+8\br_{\rm a}\br_1^2+3\br_1\br_{\rm a}^2-3\br_1^3-12\br_{\rm a}^2)\bS^6\nonumber\\
& \qquad -3\bE\br_1^3(27\br_{\rm a}^2\br_1^2-45\br_1\br_{\rm a}^2+6\br_{\rm a}\br_1+21\br_1^4-2\br_{\rm a}^2-37\br_1^3-53\br_{\rm a}\br_1^3+90\br_{\rm a}\br_1^2)\bS^5\nonumber\\
& \qquad +3\bJ\br_1^4(27\br_1\br_{\rm a}^2+21\br_1^3-48\br_{\rm a}^2+96\br_{\rm a}\br_1-53\br_{\rm a}\br_1^2-38\br_1^2)\bS^4 +3\br_1^8\bJ(-6\br_1^2+3\br_{\rm a}^2+\br_{\rm a}\br_1-12\br_{\rm a}+16\br_1)\bS^2  \nonumber\\
& \qquad -3\bE\br_1^6(6\br_{\rm a}\br_1-6\br_1^4-2\br_{\rm a}^2+17\br_1^3+3\br_1\br_{\rm a}^2+3\br_{\rm a}^2\br_1^2+\br_{\rm a}\br_1^3-18\br_{\rm a}\br_1^2)\bS^3\nonumber\\
& \qquad -\bE\br_1^9(3\br_1^4+6\br_{\rm a}^2\br_1^2-\br_1^3-3\br_1\br_{\rm a}^2+6\br_{\rm a}\br_1-2\br_{\rm a}^2-8\br_{\rm a}\br_1^3)\bS +\bJ\br_1^{10}(6\br_1\br_{\rm a}^2-6\br_{\rm a}^2-2\br_1^2+3\br_1^3-8\br_{\rm a}\br_1^2+6\br_{\rm a}\br_1) \}\ba^4\nonumber\\
& \quad + \{ \bE\br_1^2( -6\br_1^{12}+4\br_{\rm a}\br_1^{10}+4\bS^6\br_{\rm a}\br_1-42\bS^2\br_{\rm a}^2\br_1^7-126\bS^2\br_1^9-216\bS^4\br_1^5+108\bS^2\br_1^8-504\bS^4\br_{\rm a}\br_1^5+552\bS^4\br_{\rm a}\br_1^4\nonumber\\
& \qquad +144\bS^2\br_{\rm a}\br_1^8-96\bS^2\br_{\rm a}\br_1^7-12\br_{\rm a}^2\br_1^9-6\bS^6\br_1^3-2\br_{\rm a}^2\br_1^{10}+12\bS^6\br_{\rm a}\br_1^2+135\bS^4\br_{\rm a}\br_1^6-63\bS^4\br_{\rm a}^2\br_1^5\nonumber\\
& \qquad -27\bS^2\br_{\rm a}\br_1^9+9\bS^2\br_{\rm a}^2\br_1^8+27\bS^2\br_1^{10}+12\br_{\rm a}\br_1^{11}-54\bS^4\br_1^7+9\bS^6\br_{\rm a}^2\br_1^2+24\bS^6\br_{\rm a}^2+246\bS^4\br_{\rm a}^2\br_1^4-38\bS^6\br_{\rm a}^2\br_1\nonumber\\
& \qquad +198\bS^4\br_1^6-288\bS^4\br_{\rm a}^2\br_1^3 ) +\bJ\bS(\bS^2+\br_1^3)^3(-6\br_{\rm a}\br_1^2+6\br_{\rm a}\br_1-2\br_{\rm a}^2+3\br_1\br_{\rm a}^2+\br_1^3) \}\ba^3\nonumber\\
& \quad + \{ \bE\br_1^2(6\br_{\rm a}\br_1-9\br_{\rm a}\br_1^2+24\br_{\rm a}^2-21\br_1\br_{\rm a}^2+\br_1^4+3\br_1^3+4\br_1+6\br_{\rm a}^2\br_1^2-6\br_1^2)\bS^7-\bJ\br_1^2(-22\br_1\br_{\rm a}^2+24\br_{\rm a}^2+4\br_{\rm a}\br_1\nonumber\\
& \qquad +\br_1^4+6\br_{\rm a}^2\br_1^2-6\br_{\rm a}\br_1^2)\bS^6-3\bE\br_1^5(17\br_1^4+189\br_{\rm a}\br_1^2-45\br_{\rm a}\br_1^3-186\br_{\rm a}\br_1-99\br_1\br_{\rm a}^2+96\br_{\rm a}^2+24\br_{\rm a}^2\br_1^2-4\br_1\nonumber\\
& \qquad -75\br_1^3+78\br_1^2)\bS^5+3\bJ\br_1^5(-45\br_{\rm a}\br_1^3-98\br_1\br_{\rm a}^2+72\br_1^2+24\br_{\rm a}^2\br_1^2-72\br_1^3-184\br_{\rm a}\br_1+96\br_{\rm a}^2+17\br_1^4+186\br_{\rm a}\br_1^2)\bS^4\nonumber\\
& \qquad +3\bE\br_1^9(-30\br_{\rm a}-33\br_1^2+4+27\br_{\rm a}\br_1+30\br_1+10\br_1^3-9\br_{\rm a}\br_1^2+3\br_{\rm a}^2)\bS^3-3\bJ\br_1^9(2\br_{\rm a}^2-32\br_{\rm a}+36\br_1-36\br_1^2\nonumber\\
& \qquad +30\br_{\rm a}\br_1+10\br_1^3-9\br_{\rm a}\br_1^2)\bS^2-\bE\br_1^{11}(-3\br_1^3+12\br_{\rm a}^2-\br_1^4-15\br_1\br_{\rm a}^2+6\br_1^2+3\br_{\rm a}^2\br_1^2-4\br_1-6\br_{\rm a}\br_1+9\br_{\rm a}\br_1^2)\bS\nonumber\\
& \qquad +\bJ\br_1^{11}(-\br_1^4-14\br_1\br_{\rm a}^2+3\br_{\rm a}^2\br_1^2+12\br_{\rm a}^2-4\br_{\rm a}\br_1+6\br_{\rm a}\br_1^2) \}\ba^2 \nonumber\\
& \quad + \{ \br_1^3(4\br_1^{12}+2\br_{\rm a}\br_1^{12}+36\bS^2\br_{\rm a}^2\br_1^7+120\bS^2\br_1^9+144\bS^4\br_1^5-72\bS^2\br_1^8+504\bS^4\br_{\rm a}\br_1^5-360\bS^4\br_{\rm a}\br_1^4-144\bS^2\br_{\rm a}\br_1^8\nonumber\\
& \qquad +72\bS^2\br_{\rm a}\br_1^7+9\bS^2\br_1^{11}-18\bS^4\br_1^8+2\bS^6\br_{\rm a}\br_1^3+8\br_{\rm a}^2\br_1^9+4\bS^6\br_1^3-12\bS^6\br_{\rm a}\br_1^2-264\bS^4\br_{\rm a}\br_1^6+126\bS^4\br_{\rm a}^2\br_1^5\nonumber\\
& \qquad +60\bS^2\br_{\rm a}\br_1^9-18\bS^2\br_{\rm a}^2\br_1^8-54\bS^2\br_1^{10}-12\br_{\rm a}\br_1^{11}+108\bS^4\br_1^7-18\bS^6\br_{\rm a}^2\br_1^2-21\bS^4\br_{\rm a}^2\br_1^6+45\bS^4\br_{\rm a}\br_1^7+3\bS^6\br_{\rm a}^2\br_1^3\nonumber\\
& \qquad -9\bS^2\br_{\rm a}\br_1^{10}+3\bS^2\br_{\rm a}^2\br_1^9-16\bS^6\br_{\rm a}^2-252\bS^4\br_{\rm a}^2\br_1^4+36\bS^6\br_{\rm a}^2\br_1-204\bS^4\br_1^6+192\bS^4\br_{\rm a}^2\br_1^3)\bE-\br_1^3\bJ\bS(2\br_{\rm a}-3\br_{\rm a}\br_1\nonumber\\
& \qquad +\br_{\rm a}^2+3\br_1^2+4-6\br_1)(\bS^2+\br_1^3)^3 \} \ba\nonumber\\
& \quad - \br_1^3(\br_1-2)^3(-\br_{\rm a}^2\br_1^9+9\bS^2\br_1^8-9\bS^2\br_{\rm a}\br_1^7-18\bS^4\br_1^5+45\bS^4\br_{\rm a}\br_1^4-24\bS^4\br_{\rm a}^2\br_1^3+2\bS^6\br_{\rm a}^2)(\bJ-\bE\bS) \big],\\
C_1 & = \frac{(\br_{\rm a}-\br_{\rm p})}{\bar \Delta(\br_1)^2(\bS^2+\br_1^3)^2(\br_{\rm a}-\br_1)^2(\br_{\rm p}-\br_1)} \big[3\bE\bS^2\br_1(2\br_{\rm a}\bS^2+3\br_1^4-2\br_{\rm a}\br_1^3)\ba^5\nonumber\\
& \quad +(\bE\bS^4(\br_{\rm a}+\br_1+6\br_{\rm a}\br_1)-6\bJ\br_1\br_{\rm a}\bS^3+\bE\br_1^3(9\br_1^2+2\br_1+2\br_{\rm a}-3\br_{\rm a}\br_1)\bS^2+3\bJ\br_1^4(\br_{\rm a}-3\br_1)\bS+\bE\br_1^6(\br_{\rm a}+\br_1))\bS\ba^4\nonumber\\
& \quad + (2\bE\br_1^2(\br_1^6(3\br_{\rm a}-\br_1)+(9\br_1^2-20\br_1+12\br_{\rm a}-3\br_{\rm a}\br_1)\br_1^3\bS^2-\bS^4(9\br_{\rm a}+\br_1-6\br_{\rm a}\br_1)) -\bJ\bS(\bS^2+\br_1^3)^2(\br_{\rm a}+\br_1))\ba^3\nonumber\\
& \quad +(\bE(2\br_1^2+\br_1-19\br_{\rm a}+8\br_{\rm a}\br_1-4)\bS^5-2\bJ(\br_1^2-\br_1-9\br_{\rm a}+4\br_{\rm a}\br_1)\bS^4+2\bE\br_1^3(11\br_1^2-17\br_1+11\br_{\rm a}-7\br_{\rm a}\br_1-4)\bS^3\nonumber\\
& \qquad -2\bJ\br_1^3(11\br_1^2-20\br_1+12\br_{\rm a}-7\br_{\rm a}\br_1)\bS^2+\bE\br_1^6(2\br_1^2+\br_1+5\br_{\rm a}-4\br_{\rm a}\br_1-4)\bS-2\bJ\br_1^6(\br_1^2-\br_1+3\br_{\rm a}-2\br_{\rm a}\br_1))\br_1^2\ba^2\nonumber\\
& \quad + (\bE\br_1(2(\br_1^2+3\br_{\rm a}\br_1^2+8\br_{\rm a}-11\br_{\rm a}\br_1)\bS^4+\br_1^3(9\br_1^3-32\br_1^2-3\br_{\rm a}\br_1^2+16\br_{\rm a}\br_1+36\br_1-28\br_{\rm a})\bS^2+2\br_1^6(\br_1^2+\br_{\rm a}\br_1-4\br_{\rm a}))\nonumber\\
& \qquad +\bJ\bS(\bS^2+\br_1^3)^2(\br_{\rm a}-3\br_1+4))\br_1^2\ba -\br_1^3(\br_1-2)^2(\bJ-\bE\bS)(9\bS^2\br_1^4+\br_{\rm a}(4\bS^2+\br_1^3)(\bS^2-2\br_1^2)) \big]. 
\end{align}
\end{widetext}

\bibliographystyle{unsrtnat}
\bibliography{velocity}

\begin{thebibliography}{32}
\providecommand{\natexlab}[1]{#1}
\providecommand{\url}[1]{\texttt{#1}}
\expandafter\ifx\csname urlstyle\endcsname\relax
  \providecommand{\doi}[1]{doi: #1}\else
  \providecommand{\doi}{doi: \begingroup \urlstyle{rm}\Url}\fi

\bibitem[{Gillessen} et~al.(2009)]{Gillessenetal2009}
S.~{Gillessen} et~al.
\newblock {Monitoring stellar orbits around the massive black hole in the
  Galactic Center}.
\newblock \emph{Astrophys. J.}, 692:\penalty0 1075, 2009.

\bibitem[{Genzel} et~al.(2010){Genzel}, {Eisenhauer}, and
  {Gillessen}]{Genzel:etal:2010}
R.~{Genzel}, F.~{Eisenhauer}, and S.~{Gillessen}.
\newblock {The Galactic Center massive black hole and nuclear star cluster}.
\newblock \emph{Rev. Mod. Phys.}, 82:\penalty0 3121, 2010.

\bibitem[SKA()]{SKA:Web}
URL \url{http://www.skatelescope.org}.

\bibitem[Mathisson(1937)]{Mathisson:1937}
M.~Mathisson.
\newblock {Neue Mechanik materieller Systeme}.
\newblock \emph{Acta Phys. Pol.}, 6:\penalty0 163, 1937.

\bibitem[Papapetrou(1951)]{Papapetrou:1951:3}
A.~Papapetrou.
\newblock {Spinning test-particles in General Relativity. I}.
\newblock \emph{Proc. Roy. Soc. Lond. A}, 209:\penalty0 248, 1951.

\bibitem[Tulczyjew(1959)]{Tulczyjew:1959}
W.~Tulczyjew.
\newblock {Motion of multipole particles in General Relativity theory}.
\newblock \emph{Acta Phys. Pol.}, 18:\penalty0 393, 1959.

\bibitem[{Dixon}(1964)]{Dixon:1964}
W.~G. {Dixon}.
\newblock {A covariant multipole formalism for extended test bodies in General
  Relativity}.
\newblock \emph{Nuovo Cimento}, 34:\penalty0 317, 1964.

\bibitem[{Dixon}(1974)]{Dixon:1974:1}
W.~G. {Dixon}.
\newblock {Dynamics of extended bodies in General Relativity. III. Equations of
  motion}.
\newblock \emph{Phil. Trans. Roy. Soc. Lond. A}, 277:\penalty0 59, 1974.

\bibitem[{Enolskii} et~al.(2011){Enolskii}, {Hackmann}, {Kagramanova}, {Kunz},
  and {L\"ammerzahl}]{EHKKL11}
V.~Z. {Enolskii}, E.~{Hackmann}, V.~{Kagramanova}, J.~{Kunz}, and
  C.~{L\"ammerzahl}.
\newblock {Inversion of hyperelliptic integrals of arbitrary genus with
  application to particle motion in general relativity}.
\newblock \emph{J. Geom. Phys.}, 61:\penalty0 899, 2011.

\bibitem[{Enolskii} et~al.(2012)]{EHKKLS12}
V.~Z. {Enolskii} et~al.
\newblock {Inversion of a general hyperelliptic integral and particle motion in
  {H}o\v{r}ava–-{L}ifshitz black hole space--times}.
\newblock \emph{J. Math. Phys.}, 53:\penalty0 012504, 2012.

\bibitem[{Steinhoff} and {Puetzfeld}(2010)]{Steinhoff:Puetzfeld:2009:1}
J.~{Steinhoff} and D.~{Puetzfeld}.
\newblock {Multipolar equations of motion for extended test bodies in General
  Relativity}.
\newblock \emph{Phys. Rev. D}, 81:\penalty0 044019, 2010.

\bibitem[{Ehlers} and {Rudolph}(1977)]{Ehlers:Rudolph:1977}
J.~{Ehlers} and E.~{Rudolph}.
\newblock {Dynamics of extended bodies in general relativity: Center-of-mass
  description and quasirigidity}.
\newblock \emph{Gen. Rel. Grav.}, 8:\penalty0 197, 1977.

\bibitem[{Obukhov} and {Puetzfeld}(2011)]{Obukhov:Puetzfeld:2011:1}
Y.~N. {Obukhov} and D.~{Puetzfeld}.
\newblock {Dynamics of test bodies with spin in de Sitter spacetime}.
\newblock \emph{Phys. Rev. D}, 83:\penalty0 044024, 2011.

\bibitem[{Chicone} et~al.(2005){Chicone}, {Mashhoon}, and
  {Punsley}]{Chicone:2005}
C.~{Chicone}, B.~{Mashhoon}, and B.~{Punsley}.
\newblock {Relativistic motion of spinning particles in a gravitational field}.
\newblock \emph{Phys. Lett. A}, 343:\penalty0 1, 2005.

\bibitem[{Mashhoon} and {Singh}(2006)]{Mashhoon:2006}
B.~{Mashhoon} and D.~{Singh}.
\newblock {Dynamics of extended spinning masses in a gravitational field}.
\newblock \emph{Phys. Rev. D}, 74:\penalty0 124006, 2006.

\bibitem[Singh(2008{\natexlab{a}})]{Singh:2008a}
D.~Singh.
\newblock {An analytic perturbation approach for classical spinning particle
  dynamics}.
\newblock \emph{Gen. Relativ. Gravit.}, 40:\penalty0 1179, 2008{\natexlab{a}}.

\bibitem[Singh(2008{\natexlab{b}})]{Singh:2008b}
D.~Singh.
\newblock {Perturbation method for classical spinning particle motion. I. Kerr
  space-time}.
\newblock \emph{Phys. Rev. D}, 78:\penalty0 104028, 2008{\natexlab{b}}.

\bibitem[Schiff(1960)]{Schiff:1960}
L.~I. Schiff.
\newblock {Possible New Experimental Test of General Relativity Theory}.
\newblock \emph{Phys. Rev. Lett.}, 4:\penalty0 215, 1960.

\bibitem[Everitt et~al.(2011)]{GPB:2011}
C.~W.~F. Everitt et~al.
\newblock {Gravity Probe B: Final Results of a Space Experiment to Test General
  Relativity}.
\newblock \emph{Phys. Rev. Lett.}, 106:\penalty0 221101, 2011.

\bibitem[{Semer\'ak}(1999)]{Semerak:1999}
O.~{Semer\'ak}.
\newblock {Spinning test particles in a Kerr field. I}.
\newblock \emph{Mon. Not. R. Astron. Soc.}, 308:\penalty0 863, 1999.

\bibitem[{Kyrian} and {Semer\'ak}(2007)]{Semerak:2007}
K.~{Kyrian} and O.~{Semer\'ak}.
\newblock {Spinning test particles in a Kerr field. II}.
\newblock \emph{Mon. Not. R. Astron. Soc.}, 382:\penalty0 1922, 2007.

\bibitem[{Plyatsko} and {Fenyk}(2013)]{Plyatsko:2013}
R.~{Plyatsko} and M.~{Fenyk}.
\newblock {Highly relativistic circular orbits of spinning particle in the Kerr
  field}.
\newblock \emph{Phys. Rev. D}, 87:\penalty0 044019, 2013.

\bibitem[{Steinhoff} and {Puetzfeld}(2012)]{Steinhoff:Puetzfeld:2012}
J.~{Steinhoff} and D.~{Puetzfeld}.
\newblock {Influence of internal structure on the motion of test bodies in
  extreme mass ratio situations}.
\newblock \emph{Phys. Rev. D}, 86:\penalty0 044033, 2012.

\bibitem[G{arc\'{\i}a} et~al.(2013)G{arc\'{\i}a}, {Hackmann}, {Kunz},
  {L\"ammerzahl}, and {Macias}]{Garciaetal2013}
A.~G{arc\'{\i}a}, E.~{Hackmann}, J.~{Kunz}, C.~{L\"ammerzahl}, and A.~{Macias}.
\newblock {Motion of test particles in a regular black hole space--time}.
\newblock \emph{arXiv}, 1306.2549 [gr-qc], 2013.

\bibitem[{Mino}(2003)]{Mino03}
Y.~{Mino}.
\newblock {Perturbative approach to an orbital evolution around a supermassive
  black hole}.
\newblock \emph{Phys. Rev. D}, 67:\penalty0 084027, 2003.

\bibitem[Suzuki and Maeda(1998)]{SuzukiMaeda1998}
S.~Suzuki and K.~Maeda.
\newblock {Innermost stable circular orbit of a spinning particle in Kerr
  spacetime}.
\newblock \emph{Phys. Rev. D}, 58:\penalty0 023005, 1998.

\bibitem[{Tod} et~al.(1976){Tod}, {De Felice}, and {Calvani}]{TodFelice1976}
K.~P. {Tod}, F.~{De Felice}, and M.~{Calvani}.
\newblock {Spinning Test Particles in the Field of a Black Hole}.
\newblock \emph{Nuovo Cimento}, 34 B:\penalty0 365, 1976.

\bibitem[{Le Tiec} et~al.(2013)]{LeTiecetal2013}
A.~{Le Tiec} et~al.
\newblock {Periastron Advance in Spinning Black Hole Binaries: Gravitational
  Self-Force from Numerical Relativity}.
\newblock \emph{Phys. Rev. D}, 88:\penalty0 124027, 2013.

\bibitem[{Tessmer} et~al.(2013){Tessmer}, {Hartung}, and
  {Sch\"afer}]{Tessmer:etal:2013}
M.~{Tessmer}, J.~{Hartung}, and G.~{Sch\"afer}.
\newblock {Aligned spins: Orbital elements, decaying orbits, and last stable
  circular orbit to high post-Newtonian orders}.
\newblock \emph{Class. Quant. Grav.}, 30:\penalty0 015007, 2013.

\bibitem[{Rubilar} and {Eckart}(2001)]{Rubilar:Eckart:2001}
G.~{Rubilar} and A.~{Eckart}.
\newblock {Periastron shifts of stellar orbits near the Galactic Center}.
\newblock \emph{Astr. Astrophys.}, 374:\penalty0 95, 2001.

\bibitem[{Lauricella}(1893)]{Lauricella1893}
G.~{Lauricella}.
\newblock {Sulla funzioni ipergeometriche a pi\`{u} variabili}.
\newblock \emph{Rend. Circ. Math. Palermo}, 7:\penalty0 111, 1893.

\bibitem[{van Laarhoven} and {Kalker}(1988)]{vanLaarhovenetal1988}
P.~{van Laarhoven} and T.~{Kalker}.
\newblock {On the computation of Lauricella functions of the fourth kind}.
\newblock \emph{J. Comp. Appl. Math.}, 21:\penalty0 369, 1988.

\end{thebibliography}

\end{document}